\documentclass[pdflatex,sn-mathphys-num]{sn-jnl}

\usepackage{graphicx}%
\usepackage{multirow}%
\usepackage{amsmath,amssymb,amsfonts}%
\usepackage{amsthm}%
\usepackage{mathrsfs}%
\usepackage[title]{appendix}%
\usepackage{xcolor}%
\usepackage{textcomp}%
\usepackage{manyfoot}%
\usepackage{booktabs}%
\usepackage{listings}%

\usepackage{makecell}
\usepackage{url}
\usepackage{amsthm}
\usepackage{amsmath}
\usepackage{amssymb}
\usepackage{mathptmx}
\usepackage{bm}
\usepackage{amsfonts}
\usepackage[ruled,vlined]{algorithm2e}
\usepackage{natbib}

\usepackage{graphicx}
\usepackage{cuted}
\usepackage{booktabs}
\usepackage{caption}

\DeclareMathOperator{\E}{\mathbf{E}}
\DeclareMathOperator{\N}{\mathbb{N}}

\DeclareMathOperator{\U}{\mathbb{U}}
\DeclareMathOperator{\D}{\text{d}}
\DeclareMathOperator{\Var}{\mathbf{Var}}

\DeclareMathOperator{\Pj}{P_\text{jump}}

\theoremstyle{thmstyleone}%
\theoremstyle{thmstyletwo}%

\theoremstyle{thmstylethree}%

\begin{document}

\title[Efficient Mirror-type Kernels for the Metropolis-Hastings Algorithm]{Efficient Mirror-type Kernels for the Metropolis-Hastings Algorithm}


\author[1]{\fnm{Nuo} \sur{Guan}} \email{guann@mail.sustech.edu.cn}

\author*[1]{\fnm{Xiyun} \sur{Jiao}}\email{jiaoxy@sustech.edu.cn}

\affil[1]{\orgdiv{Department of Statistics and Data Science}, \orgname{Southern University of Science and Technology}, \orgaddress{\city{Shenzhen}, \postcode{518055}, \country{China}}}

\abstract{We propose a new Metropolis-Hastings (MH) kernel by introducing the Mirror move into the Metropolis adjusted Langevin algorithm (MALA). This new kernel uses the strength of one kernel to overcome the shortcoming of the other, and generates proposals that are distant from the current position, but still within the high-density region of the target distribution. The resulting algorithm can be much more efficient than both Mirror and MALA, while stays comparable in terms of computational cost. We demonstrate the advantages of the MirrorMALA kernel using a variety of one-dimensional and multi-dimensional examples. The Mirror and MirrorMALA are both special cases of the Mirror-type kernels, a new suite of efficient MH proposals. We use the Mirror-type kernels, together with a novel method of doing the whitening transformation on high-dimensional random variables, which was inspired by \citet{tan2018gaussian}, to analyse the Bayesian generalized linear mixed models (GLMMs), and obtain the per-time-unit efficiency that is 2--20 times higher than the HMC or NUTS algorithm.}




\maketitle

\section{Introduction}

Markov chain Monte Carlo (MCMC) algorithms are powerful tools to sample from complex probability distributions, and are widely used in Bayesian statistics. For a general target distribution $\begin{aligned}\pi(\theta) (\theta\in\Theta\subset\mathbb{R}^d)\end{aligned}$, which can be defined up to a constant, an MCMC algorithm produces a Markov chain $\{\theta^{(t)}\}$ whose invariant distribution is $\pi(\cdot)$. One of the most commonly used MCMC algorithms is Metropolis–Hastings \citep[MH, ][]{Metropolis1953, Hastings1970}. In each iteration, given the current state $\theta^{(t)}$, the MH algorithm draws a new state $\theta^{\prime}$ from a proposal kernel $Q$ on $\Theta$ , whose density is denoted by $q(\cdot|\theta^{(t)})$, and accept $\theta^{\prime}$ as the next state $\theta^{(t+1)}$ with the probability
\begin{equation}
\alpha(\theta^{(t)}, \theta') \triangleq 
\text{min}\left\{1, \frac{\pi(\theta')q(\theta^{(t)} | \theta')}
    {\pi(\theta^{(t)})q(\theta' | \theta^{(t)})} \right\}.
\label{eq:accept_prob}
\end{equation}
If $\theta'$ is not accepted, then $\theta^{(t+1)} = \theta^{(t)}$. The corresponding transition kernel $P$ is then 
\begin{equation}
P(\theta^{(t+1)}\mid\theta^{(t)}) = \begin{cases}
q(\theta^{(t+1)}\mid\theta^{(t)})\alpha(\theta^{(t)},\theta^{(t+1)})
&\quad\text{if }\theta^{(t+1)}\neq\theta^{(t)},
\\[1ex]
1-\int q(\theta^{(t+1)}\mid\theta^{(t)})
\alpha(\theta^{(t)},\theta^{(t+1)})\D\theta^{(t+1)}
&\quad\text{if }\theta^{(t+1)}=\theta^{(t)},
\end{cases}
\label{eq:trans_kernel}
\end{equation}
which is $\pi$-reversible since the detailed balance condition
\begin{equation}
\pi(\theta^{(t)})P(\theta^{(t+1)}\mid\theta^{(t)})=\pi(\theta^{(t+1)})P(\theta^{(t)}\mid\theta^{(t+1)})
\label{eq:detailed_balance}
\end{equation}
holds for almost each pair of $\theta^{(t)},\theta^{(t+1)}\in\Theta$. Therefore $P$ is $\pi$-invariant, and will be irreducible and aperiodic if $Q$ is irreducible and aperiodic. 

Suppose $g:\Theta\to\mathbb{R}$ is an absolutely integrable function and the expectation of $g(\theta)$ with respect to $\pi(\cdot)$ is denoted by $\pi( g) \triangleq \E _\pi[ g( \theta) ] = \int _{\Theta}g( \theta) \pi( \theta) $d$\theta$. Then the sample $\{\theta^{(t)}\}_{t=1}^T$ generated by an MCMC algorithm can be used to estimate $\pi(g)$, that is, 
\begin{equation}
\hat{\pi}(g)=\frac{1}{T}\sum_{t=1}^T g(\theta^{(t)}).
\label{eq:MC_estimate}
\end{equation}
If the Markov chain $\{\theta^{(t)}\}$ is ergodic with stationary distribution $\pi(\cdot)$, $\hat{\pi}(g)$ will converge to $\pi(g)$ almost surely as $T\to\infty$. Under some regular conditions, the central limit theorem also holds for $\sqrt{T}\hat{\pi}(g)$ with the asymptotic variance equal to
\begin{equation}
V\triangleq\lim_{T\to\infty}T\Var_\pi[\hat{\pi}(g)]
=V_g\times(1+2\sum_{k=1}^\infty\rho_k),
\label{eq:asymp_var}
\end{equation}
$V_g\triangleq\Var_\pi[g(\theta)]=
\E_\pi[g(\theta)-\pi(g)]^2$ the variance of $g(\theta)$ and $\rho_k\triangleq \text{Corr}[g(\theta^{(t)}), g(\theta^{(t+k)})]$ the lag-$k$ auto-correlation \citep{robert2004monte, ross2014introduction}. 

The efficiency of an MCMC algorithm for using $\hat{\pi}(g)$ in eq.~\ref{eq:MC_estimate} to estimate $\pi(g)$ is defined by the ratio of the true variance $V_g$ and the asymptotic variance $V$, that is,
\begin{equation}
E\triangleq\frac{V_g}V=\frac{1}{1+2\sum_{k=1}^{\infty}\rho_{k}}.
\label{eq:eff}
\end{equation}
The \emph{effective sample size} (ESS) is then equal to $TE$, which represents the number of independent samples that have the equivalent estimation power to the dependent sample $\{g(\theta^{(t)})\}_{t=1}^T$. Note that $E$ has an equivalent definition to eq.~\ref{eq:eff} as
\begin{equation}
E=\frac{V_g}{f(0)}, 
\label{eq:eff-spectral}
\end{equation}
where $f(0)$ is the spectral density of $\{g(\theta^{(t)})\}_{t=1}^T$ at frequency $0$. The spectral density $f(\omega)$ and auto-covariance $\gamma_k = V_g \rho_k$ of a stationary process are Fourier transform pairs. Their relationships are $\gamma_k = \int_{-0.5}^{0.5} e^{2\pi i\omega k} f(\omega)\D \omega$ and $f(\omega) = \sum_{k = -\infty}^{+\infty}\gamma_k e^{-2\pi i\omega k}$. 

The average acceptance probability of an MH algorithm is defined by 
\begin{equation}
\Pj=\iint\displaylimits_\Theta q(\theta'|\theta)
\alpha(\theta,\theta')\pi(\theta)\D\theta'\D\theta,
\label{eq:P_jump}
\end{equation}
which is a quantity closely related to the efficiency of the algorithm.

There are mainly two lines of efforts to improve the efficiency of the MH algorithm, which are developing new proposal kernels and optimizing proposal scales respectively. The random walk (RW) is one of the most widely used MH proposals, which generates $\theta'$ from a Gaussian or uniform distribution centred around the current state $\theta^{(t)}$. \citet{roberts1997} has proved that asymptotically, the MH algorithm with a Gaussian-RW proposal achieves the optimal efficiency if $\Pj \approx 0.234$, which implies that the scale of the RW proposal should neither be too large nor too small. The RW moves are easy to implement, but their efficiency drops quickly with the dimension $d$. For the standard-normal target distribution $\N(0, I_d)$, the optimal efficiency of the Gaussian RW is $0.233$ for $d = 1$, while $0.034$ if $d = 10$ \citep[table~1,][]{gelman1996efficient}. In fact, this decay is asymptotically of the order $d^{-1}$ \citep{roberts1997}.

Many authors have introduced new proposal kernels to avoid local or random moves. \cite{Yang2013} has proposed a novel class of ``Bactrian'' kernels, each of which is a symmetric bimodal density centred around the current state. The Bactrian kernels can generate candidates that are further away from the current value compared to RW, leading to reduced auto-correlations and improved efficiency. \cite{Jiao2025} suggested to transform the original variable to have approximately $\U(0,1)$ distribution, onto which the Bactrian proposal with reflection is applied. The reflected Bactrian proposal can lead to negative auto-correlations and super efficiency (i.e., $E>1$). Another proposal producing negative auto-correlations and super efficiency is the Mirror kernel \citep{Thawornwattana2018}, which samples new states around the ``mirror" image of the current state. The mean or mode of the target distribution is generally chosen to be the ``mirror''. 

Quite some proposals use the gradient information of the target distribution to guide the MCMC algorithms towards high-density regions. Well-known examples are Metropolis-adjusted Langevin algorithm \citep[MALA,][]{roberts1996exponential, roberts1998optimal} and Hamiltonian Monte Carlo \citep[HMC,][]{duane1987hybrid, neal2011mcmc}. MALA constructs its proposal distribution from an Euler discretization of the Langevin diffusion whose stationary distribution is the target $\pi(\cdot)$. HMC treats the original parameter $\theta$ in the model as the ``position'' variable and introduces an auxiliary ``momentum'' variable $p$ to form the complete Hamiltonian dynamics. HMC moves in the augmented $(\theta, p)$ space, with the new states generated by applying the \textit{leapfrog} integrator on the continuous-time dynamics. The Hamiltonian dynamics conserve the total energy while the integrator not. Therefore the MH acceptance/rejection is used to correct the bias. MALA is in fact a ``localised'' version of HMC, proposing the new state by applying the integrator for only one time step \citep[see, e.g.,][]{beskos2013hmc}. 

The optimal scalings of both MALA and HMC have been investigated. The acceptance probability of MALA should be tuned close to $0.574$ for optimal performance as $d\to \infty$ and the efficiency scales as $d^{-1/3}$ \citep{roberts1998optimal}. For HMC, the asymptotic optimal acceptance probability is $\Pj\approx 0.651$, with the efficiency scaling by $d^{-1/4}$. Therefore MALA and HMC are much more robust to high dimensionality compared with RW. However, it is tricky to tune the hyper-parameters of HMC manually. Thus \citet{hoffman2014no} introduced the No-U-Turn Sampler (NUTS), which automatically tunes the number of time steps and scale parameters for an HMC algorithm. NUTS was demonstrated to perform at least as efficient as a well-tuned standard HMC sampler in practice. 

\citet{girolami2011riemann} proposed the manifold-based MALA and HMC algorithms. Specifically, the parameter space of a statistical model is a Riemann manifold, since small perturbations of the parameters lead to smooth changes in the probability distribution. On the Riemann manifold, the authors defined the Langevin and Hamiltonian dynamics and discretized them to obtain the proposal distribution of the corresponding MALA and HMC algorithms. The standard MALA and HMC use the gradient information alone when exploring the parameter space, while the Riemann manifold MALA and HMC exploit the subtle geometry of the manifold, which typically leads to faster convergence, especially for high-dimensional and complex distributions. However, the Riemann manifold MALA and HMC may be time-consuming due to the additional expense in each iteration of calculating the metric tensor associated with the manifold and its derivatives.

In this article, we introduce a new kernel, named `MirrorMALA', by combining the Mirror and MALA kernels. MirrorMALA takes advantage of both Mirror and MALA, and is able to generate proposals that are distant from the current state, but still within the high-density region of the target distribution. The resulting algorithm can be much more efficient than either Mirror or MALA, while stays comparable in computational cost. We use a variety of one-dimensional and multi-dimensional examples to demonstrate the effectiveness of MirrorMALA. We then propose the Mirror-type MH kernels, of which both the Mirror and MirrorMALA proposals are special cases. We demonstrate that the Mirror-type kernels explore the parameter space in a pattern similar to HMC, while taking much less time. We further analyse the Bayesian generalized linear mixed models using the Mirror-type kernels, together with a novel method of doing whitening transformation on correlated variables, which was inspired by \citet{tan2018gaussian}, and obtain the per-time-unit efficiency that is 2-20 times higher than HMC or NUTS. 

The paper is organized as follows. In Section 2, we review the Mirror kernel,  introduce the new MirrorMALA kernel and specify the efficiency measures we use to compare different kernels. In Section 3, we use extensive one-dimensional experiments to illustrate how to tune the MirrorMALA algorithm and compare its efficiency with the RW, Mirror and MALA kernels. Section 4 shows the efficiency of MirrorMALA using multi-dimensional examples, including Gaussian targets and Bayesian logistic regression. In Section 5, we propose the Mirror-type kernels, analyse their mechanic to produce high efficiency, and use the complex GLMM examples to demonstrate the effect of Mirror-type kernels. In Section 6, we discuss the strength, weakness and applicability of Mirror-type kernels. 

\section{The New MirrorMALA Kernel}

\subsection{The Mirror kernel}
We define the Mirror kernel in contrast with RW. To sample from $\pi(\theta)$, given the current state $\theta^{(t)}$, the RW proposal samples a new state $\theta'$ by letting  
\begin{equation}
\theta^{\prime} = \theta^{(t)} +\varepsilon \sqrt{\Sigma^*} z,
\label{eq:rw_kernel}
\end{equation}
where $\Sigma^*$ is the variance-covariance matrix of $\theta$, which is typically not known exactly but estimated using the burn-in sample. The $d\times d$ matrix $\sqrt{\Sigma^*}$ is a decomposition of $\Sigma^*$ such that $\Sigma^* = \sqrt{\Sigma^*}\sqrt{\Sigma^*}^T$. We use the Cholesky decomposition and then $\sqrt{\Sigma^*}$ is a lower triangular matrix. In this paper, we add the constraint that the diagonal entries of the Cholesky factor is positive. The variable $z$ follows a standard symmetric distribution centred around $0$, for which the common choices are Gaussian and uniform. In this paper we use Gaussian distribution, that is, $z\sim \N(0, I_d)$, with $I_d$ a $d\times d$ identity matrix. Our preliminary experiments show that Gaussian and uniform RW kernels do not differ much in efficiency while Gaussian RW is more comparable to MALA. The scale parameter $\varepsilon$ should be tuned to achieve the approximately optimal $\Pj$. Since the RW proposal is symmetric around the current state $\theta^{(t)}$, the ratio $\frac{q(\theta^{(t)} | \theta')}
{q(\theta' | \theta^{(t)})} = 1$, the acceptance probability reduces to $\alpha(\theta^{(t)}, \theta') = 
\text{min}\left\{1, \frac{\pi(\theta')}{\pi(\theta^{(t)})}\right\}$, and the algorithm becomes Metropolis \citep{Metropolis1953}.

\begin{figure}[tb]
	\begin{center}
		\includegraphics[scale = 0.55]{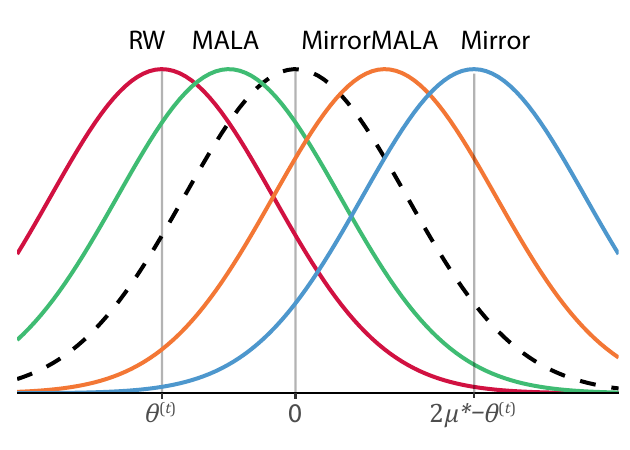}
	\end{center}
	\caption[]{The RW (red), Mirror (blue), MALA (green) and MirrorMALA (orange) proposals for sampling from the target distribution $\N(0, 1)$ (dashed black). In this plot, for both the Mirror and MirrorMALA kernels, we set $c = 1$.}
	\label{fig:normal_4_proposal}
\end{figure}

The Mirror kernel is similar to RW except that the centre of the proposal is on ``the other side'' of the target distribution, that is, 
\begin{equation}
\theta^{\prime} = \big[\mu^* + c(\mu^* - \theta^{(t)})\big] +\varepsilon \sqrt{\Sigma^*} z,
\label{eq:mirror_kernel}
\end{equation}
where $\mu^*$ is the location (mean or median) of the target distribution, generally estimated using the burn-in, and $c > 0$ is a constant. If $c = 1$, then $\mu^* + c(\mu^* - \theta^{(t)}) = 2\mu^* - \theta^{(t)}$ and the proposal centre becomes the ``mirror image'' of the current state about $\mu^*$. Note that the Mirror kernel is not Metropolis unless $c = 1$. We investigate the effects of $c$ and $\varepsilon$ on the efficiency of the Mirror kernel later in Section~3. Figure~\ref{fig:normal_4_proposal} gives the example of the RW kernel and the Mirror kernel with $c = 1$ for sampling from the $\N(0, 1)$ target.

\subsection{The MirrorMALA kernel}
We first review how the MALA kernel works. Consider the Langevin diffusion with the stationary distribution $\pi(\cdot)$ defined by the following stochastic differential equation,
\begin{equation}
\D \theta(t) = \frac{1}{2}\nabla_\theta \log \pi[\theta(t)] \D t + \D B(t),
\label{eq:mala_sde}
\end{equation}
where $B(t)$ is the Brownian motion in $\mathbb{R}^d$. A first-order Euler discretization of eq.~\ref{eq:mala_sde} gives the proposal distribution of MALA, that is,  
\begin{equation}
\theta' = \theta^{(t)}+\frac{\varepsilon^2}{2}\nabla_\theta \log \pi(\theta^{(t)}) +\varepsilon z,
\label{eq:mala_kernel}
\end{equation}
where $z\sim \N(0,I_d)$ and $\varepsilon>0$ is the integration scale parameter, which should be tuned to achieve the optimal $\Pj$. To accommodate the heterogeneity in variance of the $\theta$ components and the correlation between them, we use the strategy proposed by \citet{roberts2002langevin}, which adds a preconditioning matrix to the MALA proposal. In this paper we let this matrix be $\Sigma^*$, i.e., an estimate of the variance-covariance matrix of $\theta$. Then the MALA proposal becomes 
\begin{equation}
\theta' = \theta^{(t)} +
\frac{\varepsilon^2}{2} \Sigma^* \nabla_\theta \log \pi(\theta^{(t)}) + 
\varepsilon \sqrt{\Sigma^*} z.
\label{eq:mala_kernel_precon}
\end{equation}

The new kernel we introduce combines the Mirror kernel and the MALA kernel by using MALA to propose a new state on ``the other side'' of the target distribution. Specifically, replacing $\theta^{(t)}$ in eq.~\ref{eq:mala_kernel_precon} by $\mu^* + c(\mu^* - \theta^{(t)})$ leads to our MirrorMALA kernel, that is, 
\begin{equation}
\theta' = \mu^* + c(\mu^* - \theta^{(t)}) +
\frac{\varepsilon^2}{2} \Sigma^* \nabla_\theta \log \pi\big(\mu^* + c(\mu^* - \theta^{(t)})\big) + 
\varepsilon \sqrt{\Sigma^*} z.
\label{eq:mirrormala_kernel}
\end{equation}
We evaluate the effects of $c$ and $\varepsilon$ on the efficiency of the MirrorMALA kernel in Section 3. 

Neither the MALA nor MirrorMALA kernel is symmetric around $\theta^{(t)}$. In figure~\ref{fig:normal_4_proposal}, we show the MALA kernel and the MirrorMALA kernel with $c = 1$ for sampling from $\N(0, 1)$. Their high-density region have more overlapping with that of the target compared to RW or Mirror. 

Both the Mirror and MirrorMALA are special cases of the Mirror-type kernels, which we introduce in Section 5. 

\subsection{Illustration of the advantages of the MirrorMALA kernel}

Compared to MALA, the new MirrorMALA kernel is able to generate samples that are distant from the current state. Compared to Mirror, MirrorMALA takes into account the local geometry of the target distribution and avoids proposing states that are too far away from the high-density region (fig.~\ref{fig:normal_4_proposal}). Therefore we expect the MirrorMALA kernel to be more efficient than both component kernels. We use the following example to demonstrate its advantages. 

Suppose the target distribution $\pi(\cdot)$ is bivariate normal with mean $\mu = \left(\begin{smallmatrix}
1 \\
2
\end{smallmatrix} \right)$ and variance-covariance matrix $\Sigma = 
\left(\begin{smallmatrix}
1 & 1.8 \\
1.8 & 4
\end{smallmatrix} \right)$. To sample from it, we use each of the MH algorithms with the RW (eq.~\ref{eq:rw_kernel}), Mirror (eq.~\ref{eq:mirror_kernel}), MALA (eq.~\ref{eq:mala_kernel}) and MirrorMALA (eq.~\ref{eq:mirrormala_kernel}) proposals respectively. We first run a preliminary chain using RW with the proposal $\theta'\sim \N(0, \epsilon^2 I_2)$ to obtain estimates of the mean ($\mu$) and variance-covariance matrix ($\Sigma$), which are required when implementing all of the four algorithms. We tune the scale parameter $\epsilon$ of the RW and MALA algorithms respectively to make $\Pj \approx 0.30 $ for RW and $\Pj \approx 0.57$ for MALA. For Mirror and MirrorMALA, we set $c = 1$ and $\varepsilon = 0.5$. 

\begin{figure}[tb]
	\begin{center}
		\includegraphics[scale = 0.55]{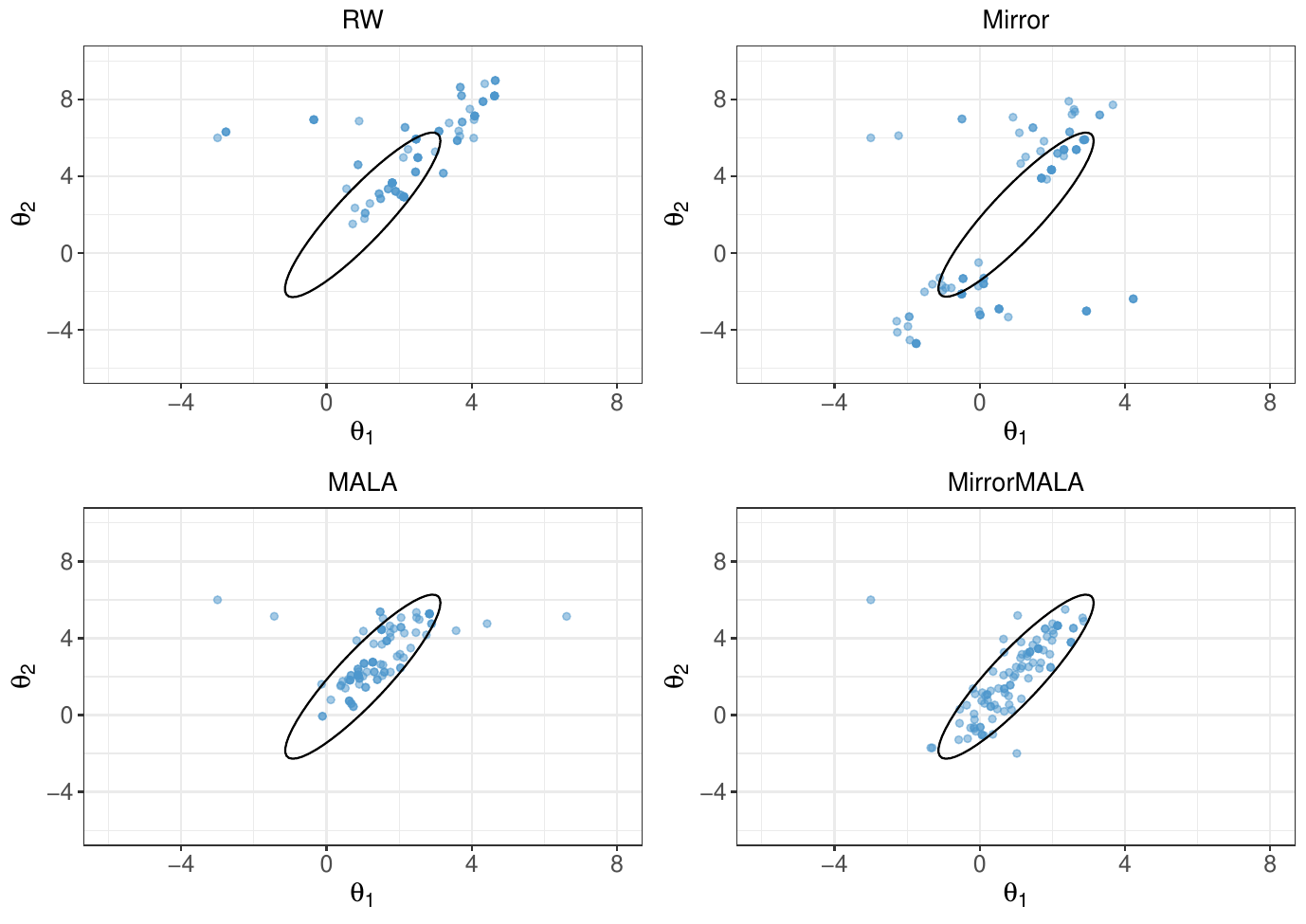}
	\end{center}
	\caption[]{The first 100 samples drawn from the bivariate normal distribution with mean $\mu = \left(\begin{smallmatrix}
		1 \\
		2
		\end{smallmatrix} \right)$ and variance-covariance matrix $\Sigma = 
		\left(\begin{smallmatrix}
		1 & 1.8 \\
		1.8 & 4
		\end{smallmatrix} \right)$ by each of the four MH algorithms with the RW, Mirror, MALA and MirrorMALA kernels respectively, starting from the identical initial point $(-3, 6)$. The elliptic shape in each subplot represents the $95\%$ contour of the target.}
	\label{fig:bivaraite_normal_4_proposal}
\end{figure}

We start all the algorithms from the same initial point $(-3, 6)$, which is far from the high-density region of the target distribution, and run 100 iterations. The sample from each algorithm are shown in figure~\ref{fig:bivaraite_normal_4_proposal}. The RW kernel mix slowly and the first 100 iterations fail to explore the whole high-density region of the target distribution. The sample drawn by the Mirror kernel scatter in a wide range. However, most of the draws do not fall into the high-density region. The MALA algorithm explores the parameter space much better than RW and Mirror, but fails to reach the high-density area that is too distant from the starting point. The MirrorMALA kernel reaches the high-density region faster than MALA, and its first 100 iterations already represents the target distribution well. 

\begin{figure}[tb] 
	\begin{center}
		\includegraphics[scale = 0.5]{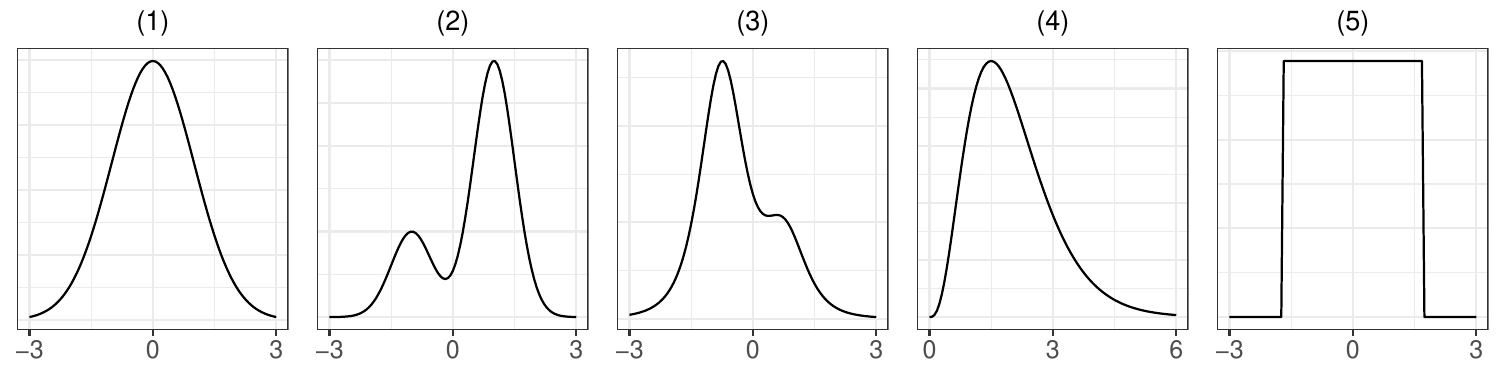}
	\end{center}
	\caption[]{Five one-dimensional target distributions: (1) $\N(0,1)$, (2) $\frac14\N(-1,\frac14)+\frac34\N(1,\frac14)$, (3) $\frac34 t_4(-\frac34,s^2)+\frac14 t_4(\frac34,s^2)$, (4) Gamma(4, 2), and (5) $\U(-\sqrt{3},\sqrt{3})$.}
	\label{fig:1_D_target}
\end{figure}

\subsection{Efficiency measures and software}

We use three efficiency measures to compare different algorithms. These measures are the lag-1 auto-correlation of the MCMC sample $\rho_1$, the average acceptance probability $\Pj$ defined in eq.~\ref{eq:P_jump} and the efficiency $E$ in eq.~\ref{eq:eff} respectively. Hereafter we focus on the efficiency of using the sample mean to estimate the true target expectation for each parameter in the distribution.

We estimate all of $\rho_1$, $\Pj$ and $E$ using the MCMC sample. Specifically, for $E$, we use the function \textsc{effectiveSize} in the R package \textsc{coda} to estimate ESS, and $E$ will be ESS dividing the total number of MCMC iterations. Note that \textsc{effectiveSize} uses the definition of $E$ in eq.~\ref{eq:eff-spectral} and calculates ESS by estimating spectral density at frequency $0$ of the Markov chain. 

For all the numerical experiments in this paper, the algorithms are programmed using R and implemented on a laptop equipped with an Intel Core i7-7500U processor (2 cores), 24 GB DDR4 2133 MHz memory, and an NVIDIA GeForce 940MX (4GB).

\section{One-dimensional Results}

In this section, we compare the efficiency of the algorithms introduced in Section~2 using the five one-dimensional (1-D) target distributions of \citet{Thawornwattana2018}, which are (1) standard normal $\N(0,1)$, with mean $0$, (2) mixture of two Gaussian distributions $\frac14\N(-1,\frac14)+\frac34\N(1,\frac14)$, with mean $\frac12$, (3) mixture of two $t_4$ distributions $\frac34 t_4(-\frac34,s^2)+\frac14 t_4(\frac34,s^2)$, where $s=\frac18\sqrt{\frac{37}2}$, with mean $-\frac38$, (4) Gamma(4, 2), with mean $2$, and (5) the uniform distribution $\U(-\sqrt{3},\sqrt{3})$, with mean $0$. The five distributions have the same variance $1$, but very different shapes (fig.~\ref{fig:1_D_target}). Both the Gamma and uniform distributions have constraints on their ranges, and for these two distributions, we apply transformations on the original variables so that the transformed variables are defined on $\mathbb{R}$. Specifically, if $\theta\sim \text{Gamma}(4, 2)$, we let $\xi=\log\theta$ and $\pi(\xi)\propto\exp(4\xi-2e^\xi)$. If $\theta\sim \U(-\sqrt{3},\sqrt{3})$,  $\xi=\log\frac{\sqrt{3}+\theta}{\sqrt{3}-\theta}$ and $\pi(\xi)\propto\frac{e^{\xi}}{\left(1+e^{\xi}\right)^2}$. We sample from $\pi(\xi)$ and get samples of $\theta$ via the inverse transformations.

\subsection{The effect of parameter $c$ on Mirror and MirrorMALA}

\begin{figure}[tb]
	\begin{center}
		\includegraphics[scale = 0.58]{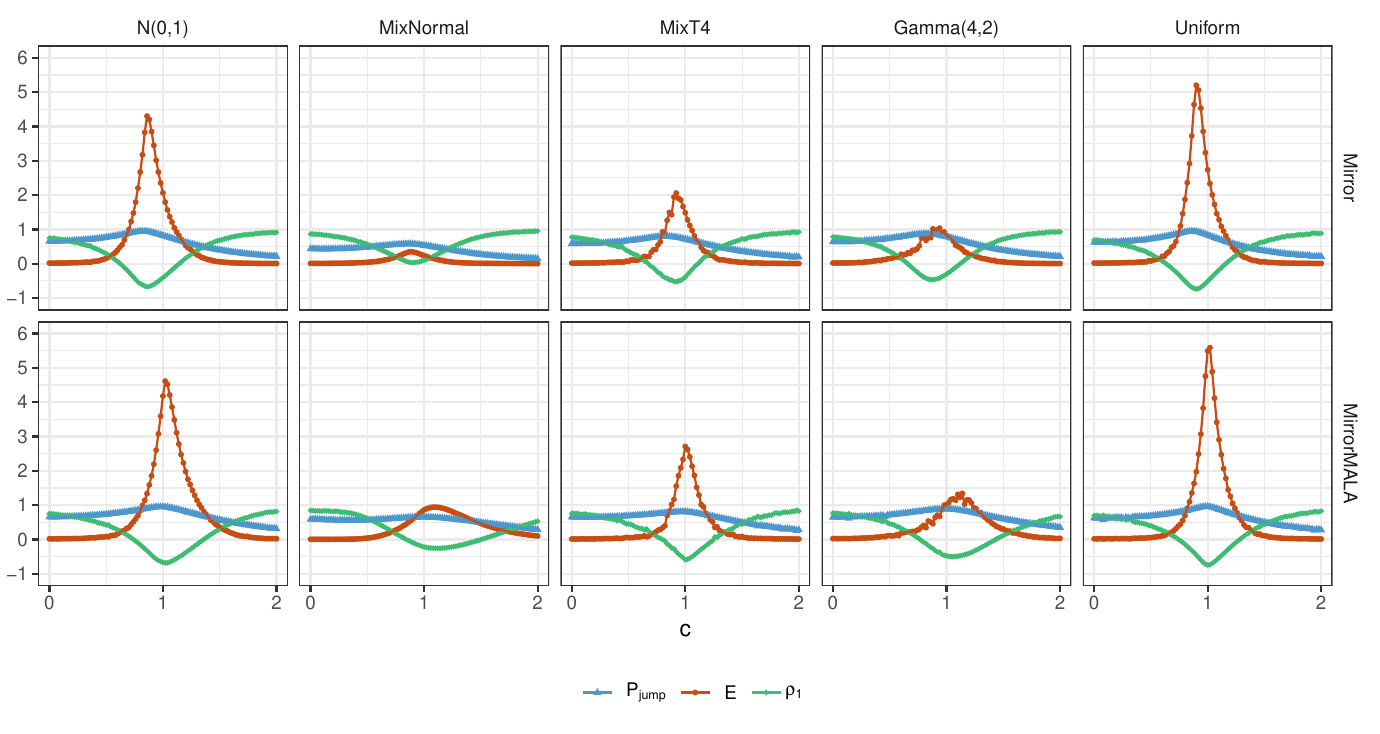}
	\end{center}
	\caption[]{The average acceptance probability $\Pj$ (blue triangle), lag-1 auto-correlation $\rho_{1}$ (green square) and efficiency $E$ (red dot) plotted against $c$ for the MALA and MirrorMALA kernels sampling from the five 1-D targets respectively. The scale parameter $\epsilon = 0.5$.}
	\label{fig:mirror-mirrormala-c}
\end{figure}

\begin{figure}[ht!]
	\begin{center}
		\includegraphics[scale = 0.7]{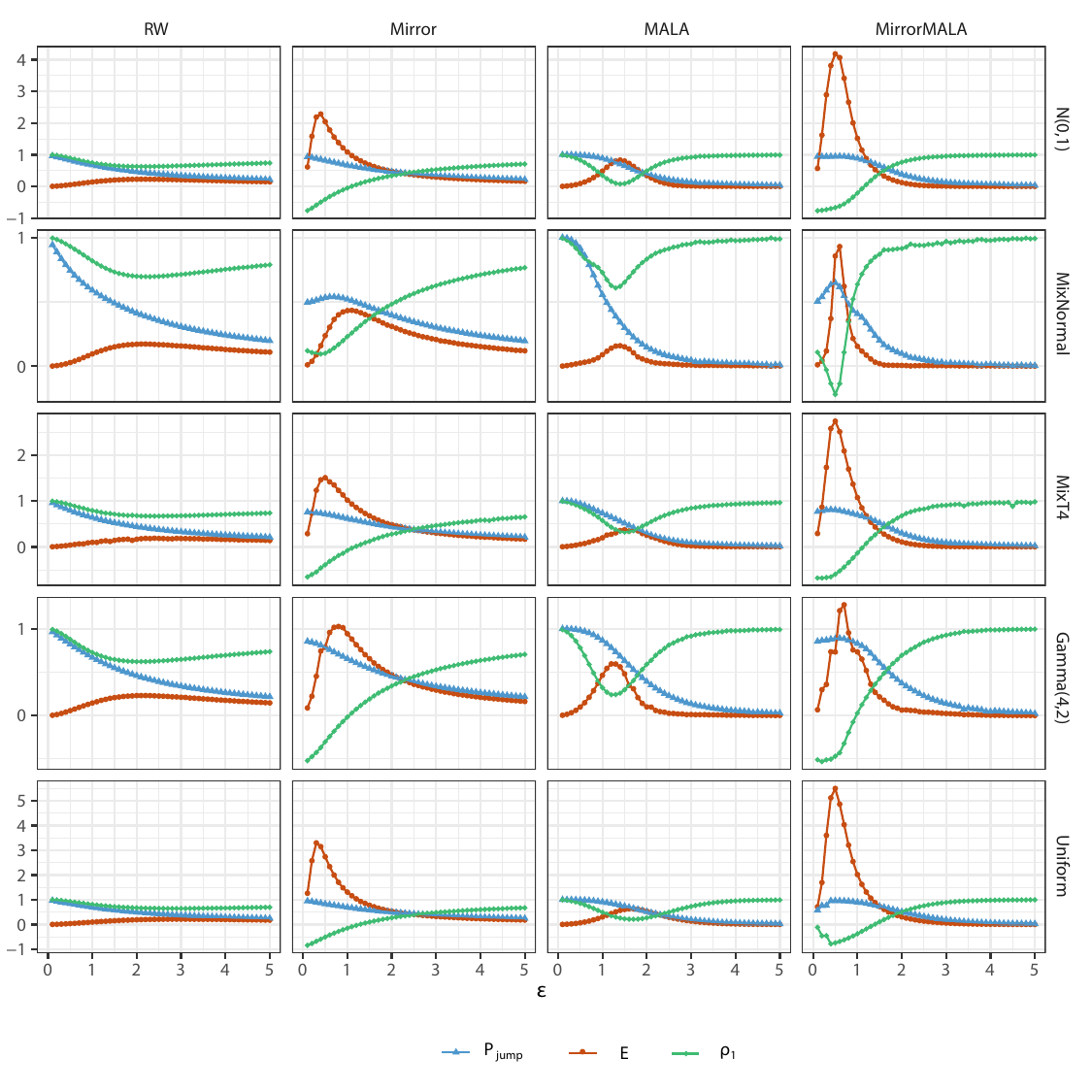}
	\end{center}
	\caption[]{The average acceptance probability $\Pj$ (blue triangle), lag-1 auto-correlation $\rho_{1}$ (green square), and efficiency $E$ (red dot) plotted against $\epsilon$ for the RW, Mirror, MALA and MirrorMALA kernels sampling from the five 1-D target distributions respectively. For the Mirror and MirrorMALA kernels, $c$ is set to $1$.}
	\label{fig:eff_compare_1_D}
\end{figure}

In this section, we investigate the effect of the $c$ parameter in the Mirror (eq.~\ref{eq:mirror_kernel}) and MirrorMALA (eq.~\ref{eq:mirrormala_kernel}) kernels on the performance of the algorithms. We fix the value of the scale parameter $\epsilon$ at $0.25$, $0.5$, $0.75$, $1$ and $2$ respectively, and sample from each of the five target distributions in figure~\ref{fig:1_D_target} using the Mirror and MirrorMALA algorithms respectively with $c$ set to each of 100 equally-spaced values between $0$ and $2$. For each target distribution, we run Mirror and MirrorMALA for $10^{6}$ iterations with a burn-in of 500 respectively. We use RW with the proposal $\theta'\sim \N(0, \epsilon^2)$ to do the burn-in sampling and utilize the 500 iterations to estimate the target mean $\mu^*$ and standard deviation $\sigma^*$.

Figure~\ref{fig:mirror-mirrormala-c} shows the average acceptance probability $\Pj$, lag-1 autocorrelation $\rho_1$ and efficiency $E$ of the Mirror and MirrorMALA algorithms as a function of $c$ for each target distribution. All of $\Pj$, $\rho_1$ and $E$ are estimated using the MCMC sample and the results are averaged over 5 replicates. In figure~\ref{fig:mirror-mirrormala-c} we set $\epsilon = 0.5$, and the results for $\epsilon = 0.25, 0.75, 1, 2$ are in figures~S1--S4 of the Supplementary Material. For all the values of $\epsilon$, $\Pj$, $\rho_1$ and $E$ show the similar trend of change with $c$ as in figure~\ref{fig:mirror-mirrormala-c}.

For each target distribution and algorithm, $\Pj$ changes with $c$ in the same pattern as $E$, while $\rho_1$ changes in the inverse pattern. Thus $\Pj$ reaches the highest and $\rho_1$ the lowest when $E$ is optimal. For both the Mirror and MirrorMALA kernels, the three efficiency measures become optimal as $c$ is close to $1$. Therefore, we set $c = 1$ for both Mirror and MirrrMALA in general.

\begin{table}[tb]
	\begin{tabular}{ccccccc}
		\toprule
		Kernel & Optimal $\varepsilon$ & $\Pj$ & $\rho_1$ & $E$ & Time(s) & $E$/Time \\
		
		\midrule
		&\multicolumn{6}{c}{Target 1: N(0,1)} \\
		RW&2.1&0.446&0.627&0.228&3.9&0.058\\
		Mirror&0.4&0.851&-0.495&2.287&4.1&0.556\\
		MALA&1.4&0.707&0.076&0.834&4.4&0.188\\
		MirrorMALA&0.5&0.948&-0.666&4.185&4.7&0.896\\
		
		&\multicolumn{6}{c}{Target 2: MixNormal} \\
		RW&2.2&0.388&0.696&0.173&4.2&0.041\\
		Mirror&1.1&0.509&0.259&0.434&4.4&0.100\\
		MALA&1.4&0.342&0.621&0.160&4.9&0.033\\
		MirrorMALA&0.6&0.617&-0.135&0.929&4.8&0.194\\
		
		&\multicolumn{6}{c}{Target 3: MixT4} \\
		RW&2.4&0.387&0.668&0.189&4.0&0.047\\
		Mirror&0.5&0.715&-0.391&1.506&4.2&0.355\\
		MALA&1.7&0.418&0.352&0.375&4.7&0.080\\
		MirrorMALA&0.5&0.804&-0.596&2.736&5.1&0.534\\
		
		&\multicolumn{6}{c}{Target 4: Gamma(4,2)} \\
		RW&2.2&0.430&0.622&0.231&4.0&0.058\\
		Mirror&0.8&0.706&-0.123&1.026&4.1&0.250\\
		MALA&1.2&0.782&0.240&0.596&4.3&0.140\\
		MirrorMALA&0.7&0.884&-0.322&1.276&4.3&0.294\\
		
		&\multicolumn{6}{c}{Target 5: Uniform} \\
		RW&3&0.374&0.646&0.214&4.0&0.054\\
		Mirror&0.3&0.893&-0.683&3.297&4.0&0.816\\
		MALA&1.7&0.649&0.209&0.634&4.4&0.146\\
		MirrorMALA&0.5&0.957&-0.742&5.499&4.4&1.256\\
		
		\bottomrule
	\end{tabular}
	\caption{The average acceptance probability ($\Pj$), lag-1 auto-correlation ($\rho_1$), efficiency ($E$), running time in seconds and the  efficiency per second corresponding to the optimal choice of the scale parameter $\epsilon$ for each of the RW, Mirrior, MALA and MirrorMALA kernels sampling from the five 1-D distributions respectively.} 
	\label{tab:eff_compare_1_D}
\end{table}

\subsection{Comparison of the efficiency of the RW, Mirror, MALA and MirrorMALA kernels}

In this section, we compare the performance of the RW, Mirror, MALA and MirrorMALA kernels for sampling from univariate targets, and also explore the impact of the scale parameter $\epsilon$ on the efficiency of these algorithms. For Mirror and MirrorMALA, we set $c = 1$. We choose 100 values between $0$ and $5$ for $\epsilon$. For each $\epsilon$ value, we sample from the five target distributions in figure~\ref{fig:1_D_target} respectively using each of the four algorithms. We run each algorithm for $10^{6}$ iterations with a burn-in of 500. The burn-in samples are generated by RW and the estimates of $\mu^*$ and $\sigma^*$ are calculated using the burn-in. Figure~\ref{fig:eff_compare_1_D} and table~\ref{tab:eff_compare_1_D} show the results, which are all averages over 5 replicates.

The Mirror-type kernels perform better than RW and MALA for all the targets, while the running time is comparable between the four algorithms (tab.~\ref{tab:eff_compare_1_D}). On average, Mirror and MirroMALA are around 5 times more efficient than MALA, and 10 times more than RW. Both Mirror and MirrorMALA can achieve `super-efficiency', that is, $E>1$, and MirrorMALA can be much more efficient than Mirror. However, the performance of Mirror and MirrorMALA depends on the choice of the scale parameter $\epsilon$. If $\epsilon$ is too far away from the optimal, neither Mirror nor MirrorMALA shows apparent advantages compared to RW or MALA. \cite{Thawornwattana2018} suggests to use $\epsilon = 1$ or $0.5$ for the Mirror kernel when sampling from a univariate target distribution. Based on figures~\ref{fig:mirror-mirrormala-c}\&\ref{fig:eff_compare_1_D} and S1--S4, we suggest to set $\epsilon = 0.5$ for both the Mirror and MirrorMALA kernels when sampling from univariate targets. Note that the RW and MALA kernels with optimal $\epsilon$s have intermediate $\Pj$ values, while the Mirror-type kernels with the optimal $\epsilon$ have high $\Pj$. For all the four algorithms, $\rho_1$ is low when $\epsilon$ is close to optimal, and for RW, MALA and MirrorMALA, optimal $\epsilon$ corresponds to the lowest $\rho_1$. 


\subsection{Analytical $\Pj$ of the RW, Mirror, MALA and MirrorMALA kernels for sampling from $\N(0,1)$}

It is generally difficult to calculate the efficiency $E$ analytically even for the 1-D targets. The measure $\Pj$ is highly correlated with $E$ (figs.~\ref{fig:mirror-mirrormala-c}\&\ref{fig:eff_compare_1_D}), while much easier to address. Therefore we investigate the deviation in the performance of the RW, Mirror, MALA and MirrorMALA kernels in terms of $\Pj$ for univariate targets. Since the posterior distribution can be approximately treated as normal if the data size is sufficiently large, we focus on the normal target. Our derivation of $\Pj$ does not depend on the mean and variance of the normal distribution. Thus without loss of generality, we set the target distribution to $\N(0,1)$ and derive $\Pj$ analytically as a function of $\epsilon$ for each of the RW, Mirror, MALA and MirrorMALA kernels. We fix the $c$ parameter for Mirror and MirrorMALA at $1$.

For sampling from $\N(0,1)$, \cite{gelman1996efficient} showed that $\Pj$ of the RM algorithm with the proposal $\theta' \sim \N(\theta^{(t)}, \epsilon^2)$ is 
\begin{equation}
\text{P}^\text{rw}_\text{jump} = \frac{2}{\pi}\tan^{-1}\left(\frac{2}{\epsilon}\right).
\label{eq:pjump_rw}
\end{equation}
The mirror algorithm has the proposal distribution $\theta' \sim \N(-\theta^{(t)}, \epsilon^2)$ and we denote its average acceptance probability by $\text{P}^\text{mirror}_\text{jump}$. \cite{Jiao2025} proved that 
\begin{equation}
\text{P}^\text{mirror}_\text{jump} = \text{P}^\text{rw}_\text{jump} = \frac{2}{\pi}\tan^{-1}\left(\frac{2}{\epsilon}\right).
\label{eq:pjump_mirror}
\end{equation}

For $\N(0,1)$, the proposals of MALA and MirrorMALA are  $\theta' \sim 
\N\left(\theta^{(t)}-\frac{\epsilon^2}{2}\theta^{(t)}, \epsilon^2\right)$ and $\theta' \sim \N\left(-\theta^{(t)}+\frac{\epsilon^2}{2}\theta^{(t)}, \epsilon^2\right)$ respectively. Then the average acceptance probability of MALA is
\begin{equation}
\begin{aligned}
\text{P}^\text{mala}_\text{jump}& = 
\frac{1}{2\pi\epsilon}\iint e^{-\frac{\left[\theta' - 
		\left(1 -\frac{\epsilon^2}{2}\right)\theta^{(t)}\right]^2}{2\epsilon^2}}
e^{-\frac{{\theta^{(t)}}^2}{2}}\min\left\{1, \frac{e^{-\frac{{\theta'}^2}{2}}
e^{-\frac{\left[\theta^{(t)} - 
		\left(1 -\frac{\epsilon^2}{2}\right)\theta'\right]^2}{2\epsilon^2}}}
{e^{-\frac{{\theta^{(t)}}^2}{2}}e^{-\frac{\left[\theta' - \left(1 -\frac{\epsilon^2}{2}\right)\theta^{(t)}\right]^2}{2\epsilon^2}}}\right\}
\D\theta'\D\theta^{(t)}\\
& = \frac{1}{2\pi\epsilon}\iint e^{-\frac{\left[\theta' - 
		\left(1 -\frac{\epsilon^2}{2}\right)\theta^{(t)}\right]^2}{2\epsilon^2}}
e^{-\frac{{\theta^{(t)}}^2}{2}}\min\left\{1, e^{-\frac{\epsilon^2}{8}\left({\theta'}^2 - {\theta^{(t)}}^2\right)}\right\}
\D\theta'\D\theta^{(t)} \\
& = \frac{1}{\pi}\left\{\cot^{-1}\left[\frac{1}{4}\epsilon(\epsilon^2 + 2)
\right] + \tan^{-1}\left(\frac{2}{\epsilon} - \frac{\epsilon}{2}\right) + \tan^{-1}\left(\frac{\epsilon}{2}\right) + \tan^{-1}\left(\frac{4\epsilon}{\epsilon^4 - 2\epsilon^2 + 8}\right)\right\}.
\end{aligned}
\label{eq:pjump_mala}
\end{equation}
By substituting $\theta^{(t)}$ in eq.~\ref{eq:pjump_mala} with $-\theta^{(t)}$, we find that the average acceptance probability of MirrorMALA is equal to that of MALA for each $\epsilon > 0$, that is, 
\begin{equation}
\text{P}^\text{mimala}_\text{jump} = \text{P}^\text{mala}_\text{jump}.
\label{eq:pjump_mirror_mala}
\end{equation} 

\begin{figure}[tb]
	\begin{center}
		\includegraphics[scale = 0.8]{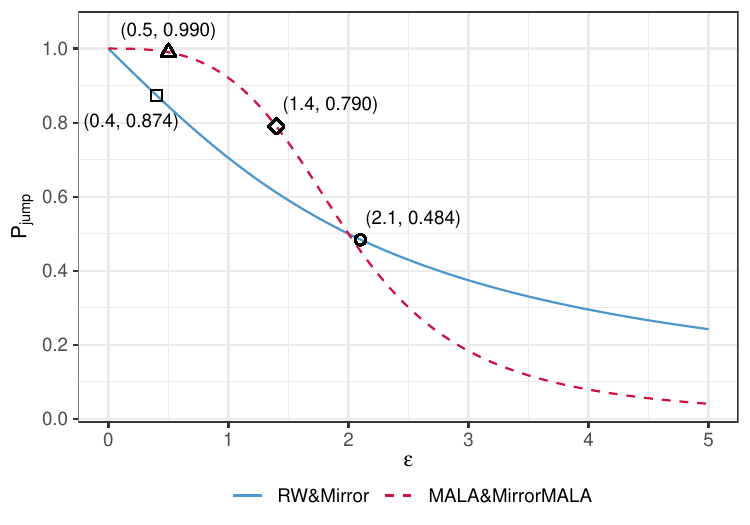}
	\end{center}
	\caption[]{The average acceptance probability $\Pj$ of the RW, Mirror, MALA and MirrorMALA algorithms for sampling from the $\N(0,1)$ target plotted against the scale parameter $\epsilon$. The $\Pj$s for RW and Mirror are equal, represented by the blue solid line. The $\Pj$s for MALA and MirrorMALA are equal, represented by the red dashed line. The symbols represent the optimal $(\epsilon,\Pj)$ for RW (circle), Mirror (square), MALA (diamond) and MirrorMALA (triangle) respectively.} 
	\label{fig:pjump}
\end{figure}

We use numerical examples to verify eqs.~\ref{eq:pjump_rw}--\ref{eq:pjump_mirror_mala}. With the optimal $\epsilon$s given in table~\ref{tab:eff_compare_1_D}, that is, $\epsilon = 2.1$ for RW, $\epsilon = 0.4$ for Mirror, $\epsilon = 1.4$ for MALA, and $\epsilon = 0.5$ for MirrorMALA, eqs.~\ref{eq:pjump_rw}--\ref{eq:pjump_mirror_mala} give $\text{P}^\text{rw}_\text{jump} = 0.484$, $\text{P}^\text{mirror}_\text{jump} = 0.874$, $\text{P}^\text{mala}_\text{jump} = 0.790$ and $\text{P}^\text{mimala}_\text{jump} = 0.990$ respectively. We then run each of the four algorithms for $10^7$ iterations with a burn-in of $10^4$ to sample from $\N(0,1)$. For each algorithm, we set the mean $\mu^*$ and standard deviation $\sigma^*$ to their true values $0$ and $1$. The estimates of $\Pj$s given by the MCMC samples are $\hat{\text{P}}^\text{rw}_\text{jump} = 0.484$, $\hat{\text{P}}^\text{mirror}_\text{jump} = 0.874$, $\hat{\text{P}}^\text{mala}_\text{jump} = 0.790$ and $\hat{\text{P}}^\text{mimala}_\text{jump} = 0.990$ respectively. The results are averages over 3 replicates, and very close to the theoretical values. The corresponding $\Pj$s in table~\ref{tab:eff_compare_1_D} are different from the values calculated here, mainly because there we use the estimates of $\mu^*$ and $\sigma^*$ obtained from a burn-in of 500 RW iterations, rather than their true values. 

We plot the $\Pj$s in eqs.~\ref{eq:pjump_rw}--\ref{eq:pjump_mirror_mala} against $\epsilon$ in figure~\ref{fig:pjump}. The $\Pj$s of all the four algorithms decrease with $\epsilon$. The $\Pj$ of MALA and MirrorMALA is higher than that of RW and Mirror for $\epsilon < 2$ but lower if $\epsilon > 2$ (The two $\Pj$s are equal at $\epsilon =2$). At optimal $\epsilon$, the Mirror and MirrorMALA algorithms have higher $\Pj$s (square and triangle in fig.~\ref{fig:pjump}) than RW and MALA, which aligns with their higher efficiency. Therefore in practice, we should tune the scale parameters of Mirror and MirrorMALA so that the average acceptance probabilities are relatively high while the efficiencies not going to infinity \citep[$E\to \infty$ as $\epsilon \to 0$,][]{Thawornwattana2018}. Our suggestion of $\epsilon = 0.5$ in Section 3.2 satisfies these requirements.  

\section{Multi-dimensional Experiments}

In this section, we compare the performance of multivariate RW, Mirror, MALA and MirrorMALA kernels using Gaussian and Bayesian logistic regression examples. 

Since all the algorithms need the estimates for the mean $\mu^*$ and variance-covariance matrix $\Sigma^*$ of the target distribution, we first investigate how the accuracy of their estimates affects the performance of the algorithms. We use each of the RW, Mirror, MALA and MirrorMALA algorithms to sample from a 2-dimensional (2-D) and 10-dimensional (10-D) Gaussian distributions respectively. The 2-D target is that in figure~\ref{fig:bivaraite_normal_4_proposal}. For the 10-D target, the mean  is sampled from $\N(0, I_{10})$, and the variance-covariance matrix is drawn from an inverse-Wishart distribution with degrees of freedom $\nu = 10$ and scale matrix $S = I_{10}$. The scale parameter $\epsilon$ for RW and MALA are tuned to optimal, while $c$ and $\epsilon$ for Mirror and MirrorMALA are set to $1$ and $0.5$ respectively. For each target, we run the RW algorithm to obtain the burn-ins of 5 different lengths ranging from $100$ to $10^5$ iterations, use each of the burn-ins to give an estimate of $\mu^*$ and $\Sigma^*$, and run all of the four algorithms with each of the estimates. The results are shown in figure~S5 of the Supplementary Material. Longer burn-in leads to higher accuracy of the estimates for $\mu^*$ and $\Sigma^*$ and improved efficiency of all the four algorithms. The Mirror and MirrorMALA algorithms always have higher efficiency than RW and MALA, and the efficiency of MirrorMALA increases much faster with the length of the burn-in than the other three algorithms. In general we suggest to use a burn-in of at least 500 iterations for low-dimensional targets with $d = 1,2$, a burn-in of at least $10^4$ for problems with $3$--$10$ dimensions, and even longer burn-in for problems with higher dimensions. 

\subsection{Multivariate Gaussian targets}

\begin{figure}[tb]
	\begin{center}
		\includegraphics[scale = 0.55]{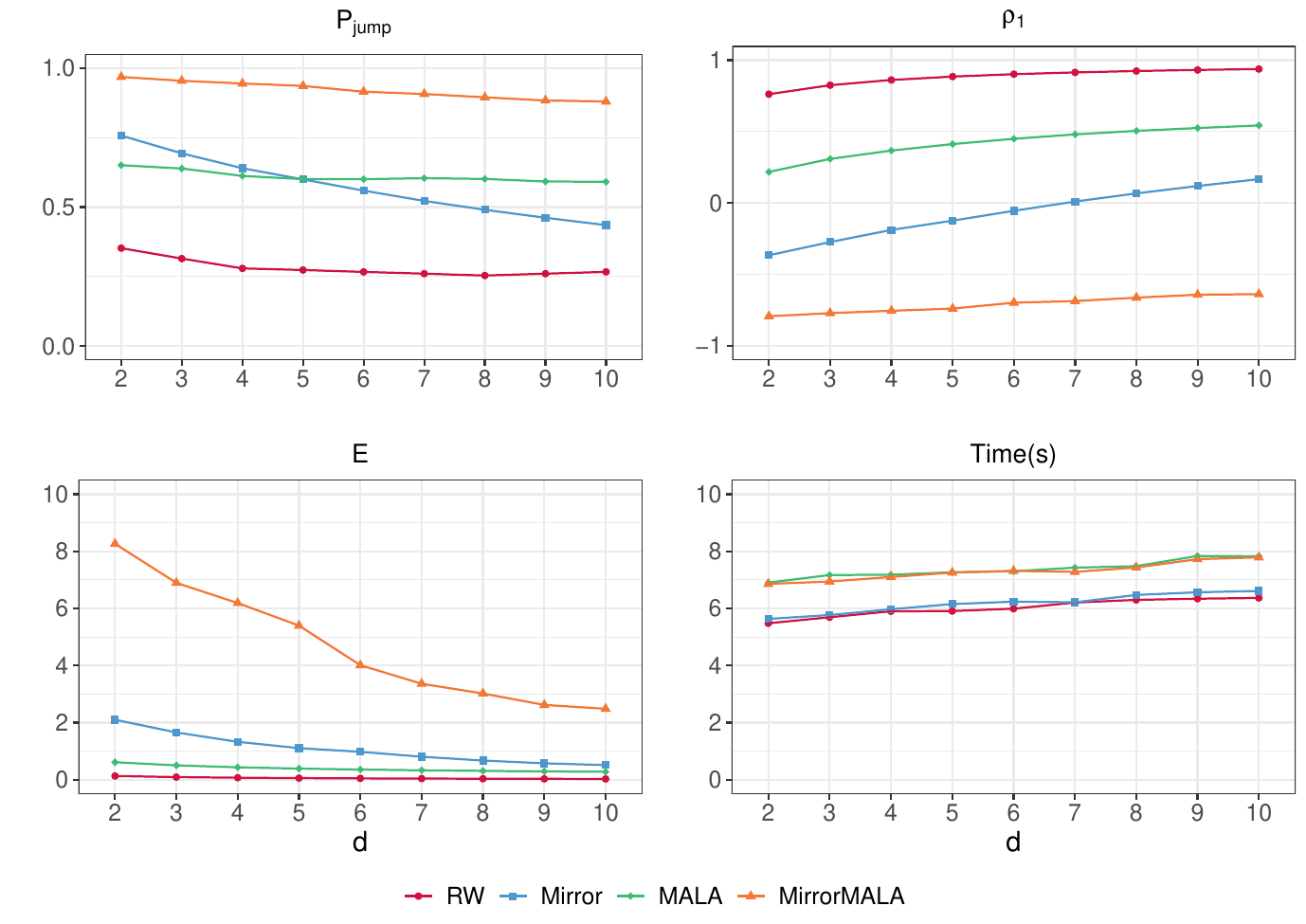}
	\end{center}
	\caption[]{The average acceptance probability ($\Pj$), lag-1 auto-correlation ($\rho_{1}$), efficiency ($E$) and running time in seconds of the RW (red), Mirror (blue), MALA (green) and MirrorMALA (orange) kernels respectively for sampling from the $\N(0, I_d)$ target distributions, plotted against the dimension $d$. Each of the $\rho_1$s and $E$s is an averages over $d$ variables.}
	\label{fig:multi-dim-gaussian}
\end{figure}

We sample from the standard Gaussian distribution $\N(0, I_d)$ with dimensions $d = 2, 3, \cdots, 10$ respectively, using the multivariate version of the RW, Mirror, MALA and MirrorMALA kernels. For each algorithm, we run a chain of $10^6$ iterations plus a burn-in of $10^4$. The burn-in samples are utilized to estimate the target mean $\mu^*$ and variance-covariance matrix $\Sigma^*$. For RW and MALA, we tune the $\epsilon$ parameter to make the $\Pj$ of RW approximately 0.234 and that of MALA around 0.574. For Mirror and MirrorMALA, we set $c = 1$, but use two $\epsilon$ values, $0.5$ and $1$. In figure~\ref{fig:multi-dim-gaussian}, we plot the $\Pj$, $\rho_1$ and $E$ and the running time (in seconds) of each algorithm against the dimension $d$. Each of the $\rho_1$s and $E$s is an average over $d$ variables and all the results are averaged over $5$ replicates. Tables~S1--S5 of the supplementary material contain the results corresponding to figure~\ref{fig:multi-dim-gaussian}. For Mirror and MirrorMALA, we plot the results for $\epsilon = 0.5$ alone in figure~\ref{fig:multi-dim-gaussian}, while including all the results for both $\epsilon = 0.5$ and $\epsilon = 1$ in tables~S1--S5.

The efficiencies of all the algorithms decrease as the dimension $d$ increases, with the $\Pj$s decreasing and $\rho_1$s increasing. However, the MirrorMALA kernel constantly performs better than the other kernels, and has efficiency higher than $2$ even for $d = 10$. MirrorMALA takes more time than RW and Mirror due to the extra step of calculating gradients, but is still the most efficient algorithm among the four even taking the computational time into account (tab.~S5). The Mirror and MirrorMALA with $\epsilon = 1$ perform worse than those with $\epsilon = 0.5$ (tabs.~S1--S5), but the MirrorMALA kernel with $\epsilon = 1$ still has higher efficiency than RW, Mirror and MALA. 

\subsection{Hundred-dimensional Gaussian distribution}

In this section, we use a 100-dimensional (100-D) Gaussian distribution $\N_{100}(0,\Sigma)$ as our target. The variance-covariance matrix $\Sigma$ is generated from an inverse-Wishart distribution with $\nu = 100$ and $S = I_{100}$. The resulting Gaussian distribution has strong correlations between the variables, with $117$ pairs of variables having absolute correlations larger than $0.9$.

\begin{table}[tb]
	\begin{tabular}{ccccccc}
		\toprule
		Kernel & Optimal $\varepsilon$ & $\Pj$ & $\rho_1$ & $E$ & Time(s) & $E$/Time$\times 10^5$\\
		
		\midrule
		RW&0.24&0.243&0.994&0.003&536&0.604\\
		Mirror1/2&0.5&0.015&0.971&0.010&566&1.819\\
		Mirror1&1&0.00002&0.99999&0.00001&563&0.001\\
		MALA&0.68&0.578&0.852&0.057&933&6.115\\
		MirrorMALA1/2&0.5&0.663&-0.240&0.323&913&35.351\\
		MirrorMALA1&1&0.082&0.871&0.020&930&2.196\\
		NUTS &Automatic&0.869&0.449&0.313&64597&0.484\\
		
		\bottomrule
	\end{tabular}
	\caption{The optimal scale ($\epsilon$), average acceptance probability ($\Pj$), lag-1 auto-correlation ($\rho_1$), efficiency ($E$), running time in seconds and efficiency per second ($\times 10^5$) for each of the RW, Mirror, MALA, MirrorMALA and NUTS algorithms sampling from the 100-dimensional Gaussian target distribution. Each of the $\rho_1$s and $E$s is an averages over the 100 variables.} 
	\label{tab:100-d-normal}
\end{table}

We use each of the RW, Mirror, MALA and MirrorMALA algorithms to sample from this Gaussian target. As in Section 4.1, for both Mirror and MirrorMALA, we set $c = 1$, but consider two $\epsilon$ values, $0.5$ and $1$. For this example, we also use the NUTS sampler implemented by the R package \textsc{RStan}, since NUTS (or HMC) is known to be efficient for sampling from high-dimensional distributions. For each of the RW, Mirror, MALA and MirrorMALA algorithms, we run a chain of $10^7$ iterations with a burn-in of $3\times10^5$ RW iterations. The burn-in samples are used to estimate the target mean $\mu^*$ and variance-covariance matrix $\Sigma^*$. Instead of utilizing all the burn-in samples for a single calculation, we estimate $\mu^*$ and $\Sigma^*$ adaptively. Specifically, each time after $5\times10^4$ runs, we compute interim $\mu^*$ and $\Sigma^*$ and employ the $\Sigma^*$ to precondition the RW algorithm for the next $5\times10^4$ runs. The estimates of $\mu^*$ and $\Sigma^*$ obtained from the last $5\times10^4$ runs are used in the main chain. As in Section 4.1, we tune $\epsilon$ to make the $\Pj$ of RW approximately 0.234 and that of MALA around 0.574, and for Mirror and MirrorMALA, we set $c = 1$, but use two $\epsilon$ values, $0.5$ and $1$. For NUTS, we set the burn-in to be $10^{5}$ and the length of the main chain $10^{7}$.

The results are shown in table~\ref{tab:100-d-normal}. Each of the $\rho_1$s and $E$s is an average over the 100 variables and all the results are averages over $5$ replicates. MirrorMALA1/2, i.e., the MirrorMALA kernel with $\epsilon = 0.5$, and NUTS perform much better than the other algorithms. However, MirrorMALA1/2 takes much less computing time than NUTS, so that MirrorMALA1/2 is the most efficient algorithm after taking the time into account, 5--70 times more efficient than the others. MirrorMALA1 performs much worse than MirrorMALA1/2, which indicates that the choice of $\epsilon$ is crucial for the efficiency of MirrorMALA when sampling from high-dimensional and complex distributions. The Mirror kernel performs poorly in this example, since most of the proposals it makes are not accepted. Therefore, for high-dimensional distributions, simply proposing distant states may not be helpful for improving the efficiency of the sampling algorithms. We need to also provide effective guidance on the traversing direction of these new states.

\subsection{Bayesian logistic regression}

In this section, we show a real-data example, which analyses the German credit dataset from the UCI repository \citep{FrankAsuncion2010} under the Bayesian logistic regression model.

\subsubsection{Model}

The German credit dataset contains the data of $N = 1,000$ customers. For customer $i$, there is one response variable, denoted by $y_i$, which is $1$ if the customer should receive credit and $0$ otherwise. There are also 24 predictors in this dataset and we denote the vector of predictors for customer $i$ by $x_i$. Then the logistic regression model is
\begin{equation}
\log\frac{p_i}{1 - p_i} = \alpha+\beta^T x_i,
\label{eq:logistic}
\end{equation}
with $y_i \sim \text{Bernoulli}(p_i)$, $\alpha$ is the intercept and $\beta$ is the 24-dimensional vector of regression coefficients, both unknown. For $\alpha$ and each component of $\beta$, we independently assign the normal prior $\N(0, \sigma^2)$ with $\sigma = 10$. Then the posterior distribution of the model is 
\begin{equation}
\begin{aligned}\pi(\alpha,\beta\mid y,x)
&\propto L(y\mid x,\alpha,\beta) \pi(\alpha) \pi(\beta)\\
&\propto\exp\left[\sum_{i=1}^N y_i(\alpha+\beta^Tx_i)
-\sum_{i=1}^N\log\left(1+\exp(\alpha+\beta^Tx_i)\right)-\frac{\alpha^2}{2\sigma^2}
-\frac{\beta^T\beta}{2\sigma^2}\right],
\end{aligned}
\label{eq:logistic-posterior}
\end{equation}
where $y = (y_1, y_2, \dots, y_N)^T$ and $x = \left(x_1, x_2, \dots, x_N\right)^T$.

\subsubsection{MCMC algorithms for the posterior distribution}

We compare the performance of RW, Mirror, MALA and MirrorMALA for sampling from the posterior distribution in eq.~\ref{eq:logistic-posterior}. For RW and MALA, we consider two versions. The first version corresponds to eqs.~\ref{eq:rw_kernel} and \ref{eq:mala_kernel}, which uses the estimated $\Sigma^*$ as the preconditioning matrix to circumvent the problem of correlations and heterogeneity of variances between variables. The second version does not use the preconditioning matrix. For Mirror and MirrorMALA, we set $c = 1$, while considering two $\epsilon$ values, $0.5$ and $1$. We also utilize the German credit dataset to assess the performance of the manifold-based MALA algorithm and NUTS. The proposal distribution of the manifold MALA algorithm is that, for the $k$th variable in the model,
\begin{equation}
\begin{array}{ll}
\theta'_{k} & = \theta_{k}^{(t)} + \frac{\epsilon^{2}}{2} 
\{ G^{-1}(\theta^{(t)}) \nabla_{\theta} \log\pi(\theta^{(t)})\}_{k} - 
\epsilon^{2} \sum_{j=1}^{d} \left\{ G^{-1}(\theta^{(t)}) 
\frac{\partial G(\theta^{(t)})}{\partial \theta_{j}} 
G^{-1}(\theta^{(t)}) \right\}_{kj} \\
& + \frac{\epsilon^{2}}{2} \sum_{j=1}^{d} \{ G^{-1}(\theta^{(t)}) \}_{kj} \operatorname{tr} \left\{ G^{-1}(\theta^{(t)}) 
\frac{\partial G(\theta^{(t)})}{\partial \theta_{j}} \right\} + 
\{ \epsilon \sqrt{G^{-1}(\theta^{(t)})} z \}_{k}
\end{array}
\label{eq:mmala_kernel}
\end{equation}
where $z\sim \N(0, I_d)$ and $G(\theta)$ is the metric tensor of the manifold. Section 7 of \citet{girolami2011riemann} gives the calculation of $G(\theta)$ and its partial derivatives for the Bayesian logistic regression model. We do not consider Riemann manifold HMC since it is much more time-consuming than the other algorithms. 

\begin{table}[tb]
	\begin{tabular}{ccccccc}
		\toprule
		Kernel & Optimal $\varepsilon$ & $\Pj$ & $\rho_1$ & $E$ & Time(s) & $E$/Time$\times 10^5$\\
		
		\midrule
		RW&0.48&0.241&0.975&0.012&1368&0.89\\
		\makecell[c]{RW \\ (no preconditioning)}&0.04&0.242&0.982&0.008&1342&0.59\\
		Mirror1/2&0.5&0.203&0.604&0.105&1401&7.49\\
		Mirror1&1&0.019&0.964&0.013&1412&0.89\\
		MALA&0.92&0.570&0.716&0.138&2278&6.08\\
		\makecell[c]{MALA \\ (no preconditioning)}&0.07&0.577&0.838&0.066&2365&2.78\\
		MirrorMALA1/2&0.5&0.728&-0.353&0.523&2922&17.89\\
		MirrorMALA1&1&0.448&0.292&0.374&2894&12.93\\
		Manifold MALA&1&0.463&0.722&0.154&92024&0.17\\
		NUTS&Automatic&0.869&-0.118&1.259&24901&5.05\\
		
		\bottomrule
	\end{tabular}
	\caption{The optimal scale parameter ($\varepsilon$), average acceptance probability ($\Pj$), lag-1 auto-correlation ($\rho_1$), efficiency ($E$), running time in seconds and efficiency per second ($\times 10^5$) for each of the RW, Mirror, MALA, MirrorMALA, manifold MALA and NUTS algorithms sampling from the posterior distribution of the Bayesian logistic regression model for the German credit dataset. Each of the $\rho_1$s and $E$s is an average over the 25 model parameters.} 
	\label{tab:logistic}
\end{table}

For each of the RW, Mirror, MALA and MirrorMALA algorithms, we set the number of burn-in to be $3\times10^4$ RW iterations and that of main chain $10^7$. For each of the algorithms requiring the estimates of $\mu^*$ or $\Sigma^*$, during the burn-in period, we update the estimates per $10^4$ iterations, and use the estimates of $\mu^*$ and $\Sigma^*$ from the last $10^4$ burn-in samples to precondition the main algorithm. For RW and MALA, we tune $\epsilon$ so that the $\Pj$ of RW is approximately 0.234 and that of MALA about 0.574. For manifold MALA, we follow \cite{girolami2011riemann} and set the scale parameter $\epsilon = 1$. We implement NUTS using \textsc{RStan} and the scale parameter is adjusted automatically. For manifold MALA and HMC, we run a chain of $10^7$ with a burn-in of $10^4$ respectively. 

The results are shown in table~\ref{tab:logistic}. Each of the $\rho_1$s and $E$s is an average over the 25 parameters in the model. MirrorMALA1/2 is much more efficient than RW, Mirror and MALA, but less efficient than NUTS. However, MirrorMALA spends much less computational time than NUTS. Therefore, MirrorMALA is the most efficient algorithm by taking the time into account, more than 3 times as efficient as NUTS. Manifold MALA is slightly more efficient than the MALA algorithm with preconditioning, but much more time-consuming, so that manifold MALA fails to show any advantages after considering the time spent. We show in figure~S6 of the Supplementary material the marginal distributions of all the parameters in this Bayesian logistic regression model, estimated using the samples of MirrorMALA1/2.


\section{Mirror-type Kernels and Their Application in Bayesian Inference of GLMMS}

The Mirror and MirrorMALA kernels are obtained by embedding the ``mirror''-reflection transformation into the RW and MALA proposal distributions respectively. In fact, we can embed the mirror reflection into any proposal MH $q(\cdot|\theta^{(t)})$. We call the class of MH transition kernels which propose around the ``mirror'' image of the current state by \emph{Mirror-type kernels}. The Mirror and MirrorMALA are both special cases of the Mirror-type kernels. 

\subsection{Understanding the high efficiency of Mirror-type kernels}

Mirror-type kernels are able to jump back and forth between the two ``sides'' of the target distribution. This may produce negative auto-correlations (on odd lags), and then lead to super efficiency according to eq.~\ref{eq:eff}. Therefore \cite{Thawornwattana2018} owed the high efficiency of the Mirror kernel mainly to the negative auto-correlations within the chain. However, we would like to emphasize another important characteristic of Mirror-type kernels, that is, they can make much larger steps than algorithms such as RW and MALA, sometimes in a pattern similar to the HMC algorithm. 

We use the bivariate-normal example in Section 2.3 to demonstrate this feature of the Mirror-type kernels. To sample from the normal distribution $\pi(\theta) =  \N\left[\mu = \left(\begin{smallmatrix}
1 \\
2
\end{smallmatrix} \right), \Sigma = \left(\begin{smallmatrix}
1 & 1.8 \\
1.8 & 4
\end{smallmatrix} \right)\right]$, we consider four algorithms, which are RW, MALA, HMC and Mirror respectively. We do not deploy the preconditioning matrix in any of the RW, MALA or Mirror algorithm. For RW and MALA, we tune $\epsilon$ to be optimal. For Mirror, we set the ``mirror'' to be the true mean $(1,2)$, $c = 1$, and $\epsilon = 0.5$. For HMC, we first introduce a bivariate normal auxiliary momentum variable $p \sim \N(0, I_2)$. Then in each iteration, we propose a new value of $(\theta, p)$ by repeating the following steps $L$ times:
\begin{equation} \nonumber
p \leftarrow p - \tfrac{\epsilon}{2} \Sigma^{-1}\theta;\,
\theta \leftarrow \theta + \epsilon p; \,
p \leftarrow p - \tfrac{\epsilon}{2} \Sigma^{-1}\theta.
\end{equation}
We manually tune $L$ and $\epsilon$ for optimal efficiency, and $L = 6$ and $\epsilon = 0.7$ are the values that we finally used.

\begin{figure}[tb]
	\begin{center}
		\includegraphics[scale = 0.55]{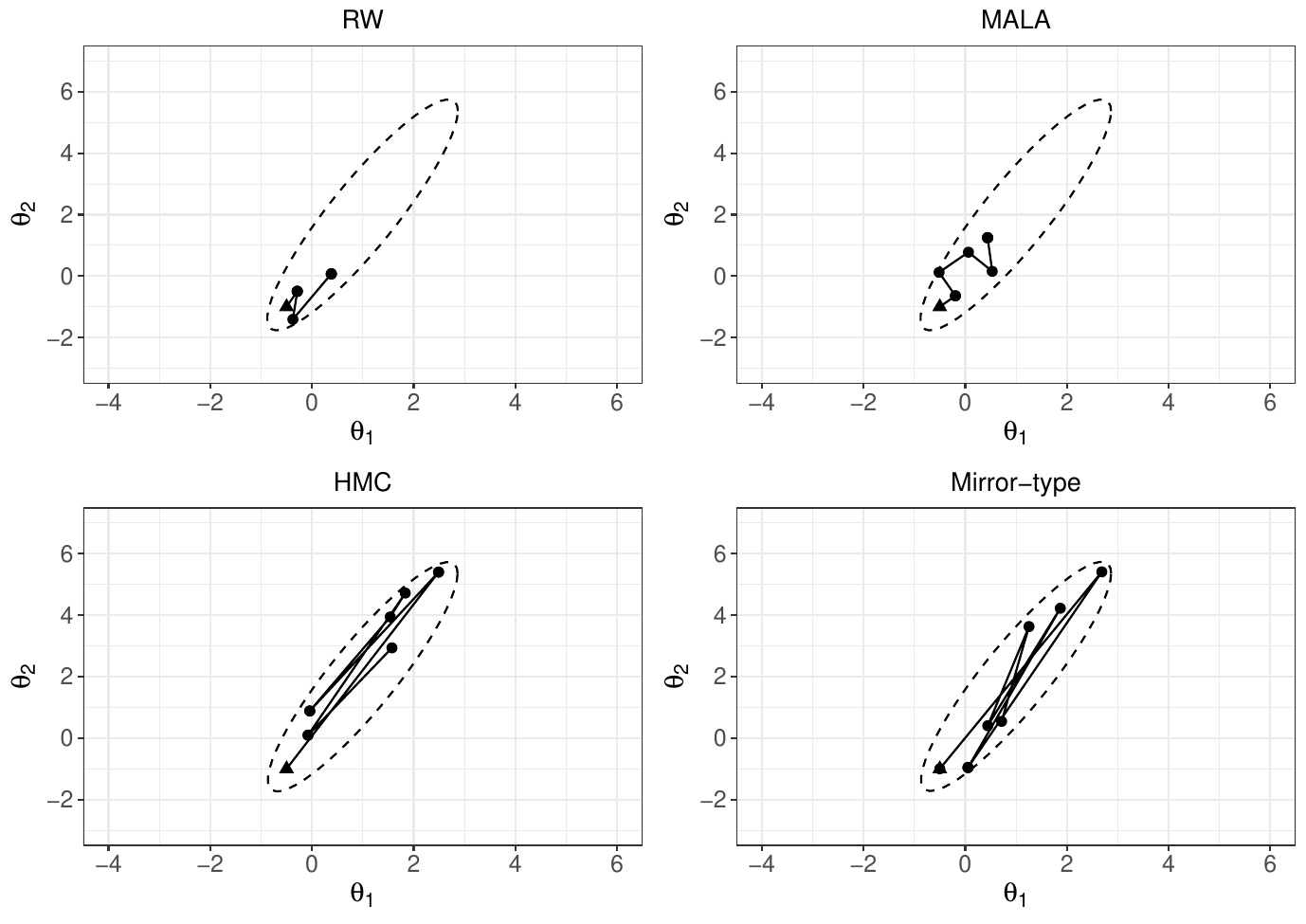}
	\end{center}
	\caption[]{The first 10 samples drawn from the target distribution $\pi(\theta) =  \N\left[\left(\begin{smallmatrix}
		1 \\
		2
		\end{smallmatrix}\right), \left(\begin{smallmatrix}
		1 & 1.8 \\
		1.8 & 4
		\end{smallmatrix} \right)\right]$ by each of the RW, MALA, HMC and Mirror algorithms, starting from the identical initial point $(-0.5, -1)$. The elliptic shape in each subplot represents the $95\%$ contour of the target.}
	\label{fig:bivaraite_normal_hmc_mirror}
\end{figure}

We start all of the four algorithms from the same point $(-0.5, -1)$. This initial point is within the 95\% contour of the target distribution, while distant from the high-density region. In figure~\ref{fig:bivaraite_normal_hmc_mirror}, we show the trajectory of the first 10 samples from each algorithm. The number of dots in each subplot corresponds to the number of accepted samples for that algorithm. For RW, $\Pj\approx 0.30$, for MALA $\Pj\approx 0.60$, for HMC $\Pj\approx 0.78$, and for Mirror $\Pj\approx 0.62$. For the first 10 steps, while RW and MALA move in a small local area around the starting point, HMC and Mirror explore the target distribution efficiently. The trajectories of HMC and Mirror display the similar moving pattern, both trying to get to the other side of the distribution. Mirror achieves this with the help of mirror reflection, while HMC via a sequence (in this example, $L = 6$) of integration steps. For this example, MirrorMALA performs similar to Mirror. Thus the Mirror-type kernels are comparable with HMC in efficiency, while much more time-saving. In fact, we can also embed the mirror reflection into the HMC proposal, and form the ``MirrorHMC'' kernel, which is another example of Mirror-type kernels. We find that MirrorHMC takes fewer integration steps to move as distant as HMC.   

As stated above, both the HMC and Mirror kernels make large steps by attempting to propose new states on ``the other side'' of the target distribution, but Mirror is much more efficient in achieving this goal. However, the efficiency of the Mirror kernel seems to largely reply on the symmetry of the target distribution due to their utility of the mirror reflection. As shown by eqs.~\ref{eq:pjump_mirror} and~\ref{eq:pjump_mirror_mala}, if the target distribution is symmetrical, proposing around the mirror image of the current state can generate distant samples without lowering the acceptance probability. On the contrary, if the target is skewed, we can imagine that the mirror proposals may hardly be accepted in many cases, which should have significant impact on the efficiency. However, the Mirror-type kernels formed by embedding the mirror reflection into gradient-based proposing mechanics, such as MALA and HMC, should be more robust to the deviation of the target distribution from symmetry, since the gradient information may help move the mirror image of the current state to regions with higher density. For the bivariate-normal example in figure~\ref{fig:bivaraite_normal_hmc_mirror}, we change the mirror from $(1,2)$ to $(0,1)$, so that the target is not symmetric around the mirror any more. Then the average acceptance probability of Mirror drops to 20\%, while that of MirrorMALA stays around 50\%. 

From the experiments in Section 4, we find that the dimension of the parameter space can have detrimental impact on the efficiency of the Mirror-type kernels, especially if there are high correlations between the variables. High dimension may also amplify the problems brought by the skewedness of the target distribution. For the logistic regression example in Section 4.3, the target distribution is just slightly skewed (fig.~S6), but already has an adverse effect on the efficiency of Mirror and MirrorMALA. To alleviate the challenges associated with high dimensionality, we recommend two techniques, which are whitening transformation and block sampling respectively. We use whitening to remove the scale heterogeneity and correlations between variables as preconditioning, but whitening may lead to computational simplicities that preconditioning cannot achieve. We use the blocking technique to divide one big task of sampling a high-dimensional variable into a few small tasks of sampling low-dimensional variables. In Section 5.2, we incorporate these two techniques into the implementation of Mirror-type kernels, and obtain per-time-unit efficiency much higher than the HMC or NUTS algorithm when sampling from the posterior distribution of Bayesian GLMMS. 

\subsection{The application of Mirror-type kernels in Bayesian GLMMs}

In this session, we propose a new whitening transformation for the parameters of the GLMMs, which preserves the conditional-independence property of the original parameters, and enables efficient block sampling. We use two real-data examples to show the effects of implementing the Mirror-type moves together with this new whitening technique and block sampling.

\subsubsection{Model}

GLMMs are extensively applied in many scientific fields, such as ecology, evolution, medicine, etc. GLMM extends the linear mixed model to accommodate non-normal response variables \citep{jiang2007linear}. We specify the GLMM in the same way as \citet{tan2018gaussian}. Suppose $y = (y_1, ..., y_n)^T$ is the matrix of all the responses, with $y_{i}=(y_{i1},\ldots,y_{in_{i}})^T$ to be the vector of the responses, and $X_{ij}$, $Z_{ij}$ to be the predictors for the $i$th subject. Then the GLMM generally has the form
\begin{equation}
\begin{aligned}g(\mu_{ij})&=X_{ij}^T\beta+Z_{ij}^T\xi_i,\text{ for }i=1,\ldots,n, j=1,\ldots,n_i,\\[2ex]\xi_i&\sim \N(0,G),\text{ for }i=1,\ldots,n,
\end{aligned}
\label{eq:glmm}
\end{equation}
where $\mu_{ij} = \E(y_{ij})$, and $g(\cdot)$ is a smooth
invertible link function. For binary responses, that is, $y_{ij}\sim\mathrm{Bernoulli}(\mu_{ij})$, $g(\mu_{ij})=\log\frac{\mu_{ij}}{1-\mu_{ij}}$, while for count responses, i.e.,  $y_{ij}\sim\mathrm{Poisson}(\mu_{ij})$, $g(\mu_{ij})=\log\mu_{ij}$. Furthermore, $\beta$ is the vector of fixed effects and $\xi_i$ the vector of random effects for the $i$th subject. As \citet{tan2018gaussian}, we denote the lower triangular matrix corresponding to the Cholesky factor of $G$ by $W$, and $W^*$ denotes the matrix for which $W_{ii}^* = \log(W_{ii})$, and $W_{ij}^* = W_{ij}$ if $i\neq j$. Then the free parameters in $G$ is denoted by $\zeta = \text{vech}(W^*)$, where $\text{vech}(W^*)$ is the vector obtained by stacking the elements of $W^*$ column by column. Suppose the prior for $(\beta, \zeta)$ is $p(\beta, \zeta) = p(\beta)p(\zeta)$. Then the posterior distribution of the GLMM in eq.~\ref{eq:glmm} is
\begin{equation}
\begin{aligned}
p(\xi_1,...,\xi_n,\beta,\zeta\mid y)&\propto 
p(y\mid \xi_1,...,\xi_n,\beta)p(\xi_1,...,\xi_n \mid \zeta)p(\beta)p(\zeta)
\\&= \prod_{i=1}^n\left\{p(\xi_i\mid\zeta)
\prod_{j=1}^{n_i}p(y_{ij}\mid\beta,\xi_i)\right\}p(\beta)p(\zeta).
\end{aligned}
\label{eq:GLMM_posterior}
\end{equation}

\subsubsection{Block sampling and new whitening transformation for Bayesian GLMMS}
We use both blocking and whitening transformation to facilitate sampling from the posterior distribution in eq.~\ref{eq:GLMM_posterior} with the MH algorithm.

We treat each $\xi_i$ as one separate block, and either sample the elements of $(\beta,\zeta)$ one by one or update $(\beta,\zeta)$ as a whole block. Suppose we treat $(\beta,\zeta)$ as one single block. Then we totally have $(n+1)$ blocks. Generally, updating by blocks may result in substantial increase of the computational time. Here we have $(n+1)$ blocks, so that the time taken by the block sampling should in theory be approximately $(n+1)$ times that of updating all the variables jointly. However, the growth of computational time is not that significant for Bayesian GLMMs, because the random effects $\xi_i$s are conditionally independent given $(\beta,\zeta)$. Specifically, given $(\beta, \zeta)$, the acceptance ratio for updating all $\xi_i$s simultaneously is 
\begin{equation}
\frac{p(\xi_1',...,\xi_n',\beta,\zeta\mid y)}{p(\xi_1,...,\xi_n,\beta,\zeta\mid y)}
= 
\frac{\prod_{i=1}^n\left\{p(\xi_i'\mid\zeta)
	\prod_{j=1}^{n_i}p(y_{ij}\mid\beta,\xi_i')\right\}}
{\prod_{i=1}^n\left\{p(\xi_i\mid\zeta)
	\prod_{j=1}^{n_i}p(y_{ij}\mid\beta,\xi_i)\right\}}.
\label{eq:full_dimensional_sampling}
\end{equation}
In contrast, if we sample $\xi_i$s as blocks, the acceptance ratio of updating each $\xi_i$ is then
\begin{equation}
\begin{aligned}\frac{p(\xi_1,...,\xi_i',...,\xi_n,\beta,\zeta\mid y)}
{p(\xi_1,...,\xi_i,...,\xi_n,\beta,\zeta\mid y)}
&= 
\frac{p(\xi_i'\mid\zeta)\prod_{j=1}^{n_i}p(y_{ij}\mid\beta,\xi_i')
\prod_{k \neq i}^n\left\{p(\xi_k\mid\zeta)
\prod_{j=1}^{n_i}p(y_{ij}\mid\beta,\xi_k)\right\}}
{p(\xi_i\mid\zeta)\prod_{j=1}^{n_i}p(y_{ij}\mid\beta,\xi_i)
	\prod_{k \neq i}^n\left\{p(\xi_k\mid\zeta)
	\prod_{j=1}^{n_i}p(y_{ij}\mid\beta,\xi_k)\right\}}
\\&=\frac{p(\xi_i'\mid\zeta)\prod_{j=1}^{n_i}p(y_{ij}\mid\beta,\xi_i')}
{p(\xi_i\mid\zeta)\prod_{j=1}^{n_i}p(y_{ij}\mid\beta,\xi_i)},
\end{aligned}
\label{eq:block_sampling}
\end{equation}
which does not depend on other random effects, so that calculating the ratio in eq.~\ref{eq:block_sampling} simply requires $1/n$ the computational effort for calculating eq.~\ref{eq:full_dimensional_sampling}. Although block sampling increases the number of updates, the reduced cost of computing the acceptance ratio offsets this, ensuring that jointly sampling all random effects and updating them separately lead to comparable computational time. Since the MH algorithm generally performs much better for updating low-dimensional variables than updating high-dimensional ones, treating $\xi_i$s as blocks should be a better strategy than sampling all of them jointly for Bayesian GLMMS.

Let $\eta=(\beta,\zeta)$ and $\theta=(\xi_1,...,\xi_n,\eta)^T$. To account for the heterogeneity in scales and correlations between variables, we perform whitening transformation on $\theta$. Suppose $\Sigma^*$ is the estimated variance-covariance matrix of $\theta$, and $C =\sqrt{\Sigma^*}$ is the Cholesky factor. The standard way to do whitening transformation on $\theta$ is to use $C^{-1}$, i.e., 
\begin{equation}
\phi=C^{-1}\theta.
\label{eq:std_transform}
\end{equation} 
We partition both $\phi$ and $C$ with respect to $\theta=(\xi_1,...,\xi_n,\eta)^T$. Then
\begin{equation}
\theta  = 
\begin{pmatrix}
\xi_1\\\xi_2\\\vdots\\\xi_n\\\eta
\end{pmatrix} 
=C\phi
= \begin{pmatrix}
C_{11}&0&\cdots&0&0\\
C_{21}&C_{22}&\cdots&0&0\\
\vdots&\vdots&\ddots&\vdots&\vdots\\
C_{n1}&C_{n2}&\cdots&C_{nn}&0\\
C_{n+1,1}&C_{n+1,2}&\cdots&C_{n+1,n}&C_{n+1,n+1}
\end{pmatrix}
\begin{pmatrix}
\phi_1\\\phi_2\\\vdots\\\phi_n\\\phi_{n+1}
\end{pmatrix}.
\label{eq:dense_whitening}
\end{equation}
Substituting $\theta$ by $C\phi$ in eq.~\ref{eq:GLMM_posterior}, we obtain the posterior distribution of $\phi$. However, $\phi_i$ $(i=1, 2,\dots, n)$ are not conditionally independent any more. Thus if we update $\phi$ by blocks, the acceptance ratios are not as simple as eq.~\ref{eq:block_sampling}, so that the computational time may largely increase.

\begin{algorithm}[tb]
	\caption{MCMC for Bayesian GLMMs with whitening \& block sampling}
	\KwIn{target $\pi(\theta)$, initial values $\theta_0$, number of burn-in $B$, number of iterations $T$, whitening flag $I_\text{sparse}$}
	\KwOut{samples $\{\theta^{(1)}, \dots, \theta^{(T)}\}$ from $\pi(\theta)$}
	
	Initialize $\theta^{(0)} \leftarrow \theta_0$ \\
	
	\For{$t \leftarrow 1$ \KwTo $B$}{
		sample $\theta^* \sim q_\text{rw}(\cdot|\theta^{(t-1)})$;
		calculate $\alpha \leftarrow \min\left\{1, \frac{\pi(\theta^*)}
		{\pi(\theta^{(t-1)})}\right\}$; \\	
		sample $u \sim \text{Uniform}(0,1)$; if $u \leq \alpha$, $\theta^{(t)} \leftarrow \theta^*$, else $\theta^{(t)} \leftarrow \theta^{(t-1)}$\\
	}
    Collect the burn-in samples $\{\theta^{(1)}, \dots, \theta^{(B)}\}$; use the burn-in to estimate $\mu^*$ and $\Sigma^*$\\
    \If{$I_{\rm sparse} = 1$}{
    	Let $\Omega^* = (\Sigma^*)^{-1}$, $R = \sqrt{\Omega^*}^T$, 
    	partition $R$, constrain some entries to be 0 as eq.~\ref{eq:sparse_whitening} \\
    	Let $\psi = R\theta$, $\pi(\psi) = |R|^{-1}\pi(R^{-1}\psi)$, and $\mu_\psi = R\mu^*$\\
    	Initialize $\theta^{(0)} \leftarrow \theta^{(B)}$; let $\psi^{(0)} = R\theta^{(0)}$\\
    	\For{$t \leftarrow 1$ \KwTo $T$}{
    	   $\psi^* \leftarrow \psi^{(t-1)}$, $\psi^{(t)} \leftarrow \psi^{(t-1)}$\\
    	   \For{$i \leftarrow 1$ \KwTo $n + 1$}{
    	   	  sample $\psi_i^* \sim q(\cdot|\psi_i^{(t)})$;
    	   	  calculate $\alpha \leftarrow \min\left\{1, \frac{\pi(\psi^*)q(\psi_i^{(t)}|\psi_i^*)}{\pi(\psi^{(t)})q(\psi_i^*|\psi_i^{(t)})}\right\}$; \\	   	  
    	   	  sample $u \sim \text{Uniform}(0,1)$; if $u \leq \alpha$, $\psi_i^{(t)} \leftarrow \psi_i^*$, else $\psi_i^* \leftarrow \psi_i^{(t)}$\\
    	   }
        Let $\theta^{(t)} = R^{-1}\psi^{(t)}$	
    	}
    }
	\Else{
		Let $C = \sqrt{\Sigma^*}$; partition $C$ w.r.t. the blocks of $\theta$\\
		Let $\phi = C^{-1}\theta$, $\pi(\phi) = |C|\pi(C\phi)$ and $\mu_\phi = C^{-1}\mu^*$\\
		Initialize $\theta^{(0)} \leftarrow \theta^{(B)}$; let $\phi^{(0)} = C^{-1}\theta^{(0)}$\\
		The updates of $\phi$ are similar to $\psi$ above, omitted here
	}
	\Return{$\{\theta^{(1)}, \dots, \theta^{(T)}\}$}
	\label{alg:block_whitening}
\end{algorithm}

We call $C^{-1}$ the dense whitening matrix. We now introduce a new sparse whitening matrix for Bayesian GLMMs, which preserves the conditional independence between $\xi_i$s. The idea is inspired by \citet{tan2018gaussian}, where the authors used Gaussian variational inference (VI) to approximate the posterior distribution of the GLMM in eq.~\ref{eq:GLMM_posterior}. Considering the conditional independence of $\xi_i$s given $\eta$, \citet{tan2018gaussian} imposed sparsity on the precision matrix of the Gaussian approximating distribution, which leads to efficient VI algorithms. Here we take advantage of the same sparse structure when specifying the whitening matrix for GLMMs. For multivariate Gaussian distributions, if the $i$th and $j$th variables are conditionally independent given the other variables, the corresponding off-diagonal elements of the precision matrix are $0$. Although the posterior distribution in eq.~\ref{eq:GLMM_posterior} is not Gaussian, we presume it to be close as long as the data size is not too small. Suppose $\Omega^* = (\Sigma^*)^{-1}$ is the estimated precision matrix of $\theta$, and we specify the following sparse structure,
\begin{equation}
\Omega^*=\begin{pmatrix}
\Omega^*_{11}&0&\cdots&0&\Omega^*_{1,n+1}\\
0&\Omega^*_{22}&\cdots&0&\Omega^*_{2,n+1}\\
\vdots&\vdots&\ddots&\vdots&\vdots\\
0&0&\cdots&\Omega^*_{nn}&\Omega^*_{n,n+1}\\
\Omega^*_{n+1,1}&\Omega^*_{n+1,2}&\cdots&\Omega^*_{n+1,n}&\Omega^*_{n+1,n+1}
\end{pmatrix},
\label{eq:precision}
\end{equation}
where $\Omega^*$ is also partitioned according to $\theta$, and $\Omega^*_{ij} = 0$ for $i \neq j$ and $i,j=1,...,n$. Suppose $R = {\sqrt{\Omega^*}}^T$. Then  
\begin{equation}
R= \begin{pmatrix}
R_{11}&0&\cdots&0&R_{1,n+1}\\
0&R_{22}&\cdots&0&R_{2,n+1}\\
\vdots&\vdots&\ddots&\vdots&\vdots\\
0&0&\cdots&R_{nn}&R_{n,n+1}\\
0&0&\cdots&0&R_{n+1,n+1}
\end{pmatrix},
\label{eq:sparse_whitening}
\end{equation}
which is a sparse upper triangular matrix. $R$ is the new whitening matrix we specify for $\theta$. Let 
\begin{equation}
\psi=R\theta,
\label{eq:new_transform}
\end{equation}
and we have 
\begin{equation}
\theta = \begin{pmatrix}
\xi_1\\\xi_2\\\vdots\\\xi_n\\\eta
\end{pmatrix} =
R^{-1}\psi=
{\begin{pmatrix}
R_{11}&0&\cdots&0&R_{1,n+1}\\
0&R_{22}&\cdots&0&R_{2,n+1}\\
\vdots&\vdots&\ddots&\vdots&\vdots\\
0&0&\cdots&R_{nn}&R_{n,n+1}\\
0&0&\cdots&0&R_{n+1,n+1}
\end{pmatrix}}^{-1}
\begin{pmatrix}
\psi_1\\\psi_2\\\vdots\\\psi_n\\\psi_{n+1}
\end{pmatrix}.
\label{eq:sparse_transform}
\end{equation}
Since $R^{-1}$ has the same structure as $R$, eq.~\ref{eq:sparse_transform} shows that the transformed variable $\psi$ keeps the conditionally independent structure of $\theta$, so that the computational cost of updating $\psi$ by blocks is similar to updating $\theta$ by blocks, both much more efficient than updating $\phi$ in eq.~\ref{eq:std_transform} by blocks. 

Algorithm~\ref{alg:block_whitening} displays the pseudo code of the MCMC algorithm for sampling from the posterior distribution of the Bayesian GLMM in eq.~\ref{eq:GLMM_posterior} using both whitening and blocking.

\subsubsection{Real-data analyses}

\paragraph{Epilepsy data}

The epilepsy data \citep{thall1990some} were collected from a clinical trial involving 59 epileptics. Each patient was randomly assigned to receive either a new drug ($\mathrm{Trt} = 1$) or a placebo ($\mathrm{Trt} = 0$). The response
is the number of seizures patients had during four follow-up periods. Other covariates include the logarithm of 1/4 the number of baseline seizures (Base), the logarithm of age (Age), which we center using the mean age of all the patients, and a binary variable V4 which is 1 for the fourth visit and 0 otherwise. We use the Poisson mixed-effects model proposed by \cite{breslow1993approximate} to analyse this dataset, that is, 
\begin{equation}
\log\mu_{ij}=\beta_0+\beta_1{\text{Base}_i}+\beta_2{\text{Trt}_i}+\beta_3{\text{Age}_i}+\beta_4\text{Base}_i\times\text{Trt}_i+\beta_5{\text{V}4_{ij}}+\xi_i
\label{eq:epilepsy}
\end{equation}
for $i = 1,\dots,59$, $j = 1,2,3,4$. The random effect $\xi_i\sim \N\left(0,\exp(2\zeta)\right)$, and we specify the following priors, $\beta_k \sim \N(0,100)$ for $k = 0, 1, \dots, 5$ and $\zeta\sim \N(0,100)$.

\begin{table}[tb]
	\begin{tabular}{lcccccc}
		\toprule
		Kernel & Optimal $\varepsilon$ & $\Pj$ & $\rho_1$ & $E$ & Time (s) & $E$/Time$\times 10^5$\\
		
		\midrule
		RW&0.29&0.235&0.990&0.005&303&1.58\\
		RW (sparse)&2.53&0.427&0.642&0.21&2635&7.97\\	
		MALA&0.76&0.576&0.804&0.094&774&12.20\\
		MALA (sparse)&1.56&0.714&0.145&0.723&6463&11.19\\
		\makecell[l]{Mirror (dense)}
		&0.5&0.764&-0.179&0.977&13196&7.41\\
		\makecell[l]{Mirror (sparse)}
		&0.5&0.820&-0.294&1.123&2872&39.11\\
		\makecell[l]{MirrorMALA (dense)}
		&0.5&0.836&-0.273&1.257&70684&1.78\\
		\makecell[l]{MirrorMALA (sparse)}
		&0.5&0.921&-0.456&1.643&7118&23.08\\
		HMC &Automatic&0.972&-0.092&1.031&62138&1.66\\
		NUTS &Automatic&0.935&0.029&0.728&59103&1.23\\
		
		\bottomrule
	\end{tabular}
	\caption{The optimal scale ($\varepsilon$), average acceptance probability ($\Pj$), lag-1 auto-correlation ($\rho_1$), efficiency ($E$), running time in seconds and efficiency per second ($\times 10^5$) for each of the 10 algorithms sampling from the posterior distribution of the GLMM for the epilepsy data. Each of the $\rho_1$s and $E$s is an average over the 66 model parameters.} 
	\label{tab:epilepsy}
\end{table}

For the GLMM in eq.~\ref{eq:epilepsy}, $\theta = (\xi_1, \dots, \xi_{59}, \beta_0, \dots, \beta_5, \zeta)^T$. There are 66 unknown parameters in total and we treat them as 60 blocks, i.e., each $\xi_i$ a block and $(\beta, \zeta)$ a block. To sample from the posterior distribution of $\theta$, we consider the following 10 algorithms.  
\begin{itemize}
	\item RW: use a 66-dimensional (66-D) RW move as in eq.~\ref{eq:rw_kernel} to update $\theta$. 
	\item RW (sparse): Algorithm~\ref{alg:block_whitening} with $I_\text{sparse} = 1$, and the proposals $q(\cdot|\psi_i^{(t)})$ are RW. Note that for the block of $(\beta, \zeta)$, we update the $7$ components iteratively using 1-D RW.
	\item MALA: use a 66-D MALA move as in eq.~\ref{eq:mala_kernel} to update $\theta$.
	\item MALA (sparse): Algorithm~\ref{alg:block_whitening} with $I_\text{sparse} = 1$, and the proposals $q(\cdot|\psi_i^{(t)})$ are MALA. For $(\beta, \zeta)$, we update the $7$ components iteratively using 1-D MALA.
	\item Mirror (dense): Algorithm~\ref{alg:block_whitening} with $I_\text{sparse} = 0$, and the proposals $q(\cdot|\phi_i^{(t)})$ are Mirror. For $(\beta, \zeta)$, we update the $7$ components iteratively using 1-D Mirror.
	\item Mirror (sparse): Algorithm~\ref{alg:block_whitening} with $I_\text{sparse} = 1$, and the proposals $q(\cdot|\psi_i^{(t)})$ are Mirror. For $(\beta, \zeta)$, we update the $7$ components iteratively using 1-D Mirror.
	\item MirrorMALA (dense): Algorithm~\ref{alg:block_whitening} with $I_\text{sparse} = 0$, and the proposals $q(\cdot|\phi_i^{(t)})$ are MirrorMALA. For $(\beta, \zeta)$, we update the $7$ components iteratively using 1-D MirrorMALA.
	\item MirrorMALA (sparse): Algorithm~\ref{alg:block_whitening} with $I_\text{sparse} = 1$, and the proposals $q(\cdot|\psi_i^{(t)})$ are MirrorMALA. For $(\beta, \zeta)$, we update the $7$ components iteratively using 1-D MirrorMALA.
	\item HMC: the static HMC algorithm implemented by the \textsc{RStan} package, constraining the integration time $L\epsilon$ to be constant.
	\item NUTS: the NUTS implemented by \textsc{RStan}, automatically tuning the integration time.
\end{itemize} 
For all of the RW, MALA, Mirror and MirrorMALA algorithms, we set the number of burn-in to be $3\times10^5$, and the number of main chain $10^7$. We use the burn-in samples to estimate the target mean $\mu^*$ and variance-covariance matrix $\Sigma^*$ adaptively as the logistic regression example in Session 4.3. For HMC and NUTS, we run a chain of $10^7$ with a burn-in of $10^5$. The scale parameter $\epsilon$ for RW and MALA is tuned to be optimal while that for HMC and NUTS is adjusted automatically. For Mirror and MirrorMALA, we set $\epsilon = 0.5$. 

The results are shown in table~\ref{tab:epilepsy}. The Mirror-type kernels perform much better than other algorithms in terms of efficiency. The Mirror and MirrorMALA with the sparse whitening are more efficient than the Mirror and MirrorMALA with the dense whitening, while taking a factor of 4-9 less time. The HMC and NUTS are comparable to the Mirror-type kernels in efficiency, but much more time-consuming. Therefore the Mirror-type kernels are 2--30 times better than the other algorithms with regard to per-time-unit efficiency.  

\paragraph{Polypharmacy data}

The polypharmacy dataset \citep{hosmer2013applied} consists of the data for 500 subjects, who were observed over seven years. The response is a binary variable that is 1 if the subject is taking drugs from three or more different groups, and 0 otherwise. The covariates include $\text{Gender} = 1$ if male and 0 if female, $\text{Race} = 0$ if white and 1 otherwise, Age, the binary
indicators for the number of outpatient mental health visits, which are $\text{MHV}\_1 = 1$ if $1\leq\text{MHV}\leq5$, $\text{MHV}\_2 = 1$ if $6\leq\text{MHV}\leq14$ and $\text{MHV}\_3 = 1$ if $\text{MHV}\geq15$, and the indicator $\text{INPTMHV} = 0$ if there were no inpatient mental health visits and 1 otherwise.  We use the logistic mixed model considered by \citep{hosmer2013applied}, that is, 
\begin{equation}
\begin{aligned}
\operatorname{logit}(\mu_{ij})& \begin{aligned}&=\beta_0+\beta_1\text{Gender}_i+\beta_2\text{Race}_i+\beta_3\text{Age}_{ij}\end{aligned}  \\
&+\beta_4\text{MHV}\_1_{ij}+\beta_5\text{MHV}\_2_{ij}+\beta_6\text{MHV}\_3_{ij} \\
&+\beta_7\text{INPTMHV}_{ij}+\xi_i,
\end{aligned}
\label{eq:polypharmacy}
\end{equation}
for $i = 1,\dots,500$, $j = 1,\dots,7$. The random intercept $\xi_i\sim \N\left(0,\exp(2\zeta)\right)$, and we use the priors $\beta_k \sim \N(0,100)$ for $k = 0, 1, \dots, 7$ and $\zeta\sim \N(0,100)$.

There are totally 509 unknown parameters in the GLMM of eq.~\ref{eq:polypharmacy}. Considering both time and efficiency, we choose to use the RW, MALA, Mirror (sparse), MirrorMALA (sparse) and NUTS described above to sample from the posterior distribution of eq.~\ref{eq:polypharmacy}. For each sampler, we run a main chain of $10^6$ iterations. For NUTS, we set the burn-in to be $10^5$. For other algorithms, we use the Gaussian variational approximation approach in \cite{tan2018gaussian} to estimate the required mean and sparse precision matrix, so that the computational time spent is much less than running a long burn-in. The results are shown in table~\ref{tab:polypharmacy}. All of Mirror (sparse), MirrorMALA (sparse) and NUTS show high efficiency, while Mirror and MirrorMALA with sparse whitening takes much less time. In fact, for this example, Mirror (dense) and MirrorMALA (dense) are about 55 and 115 times slower than Mirror (sparse) and MirrorMALA (sparse), while not showing any improvement in efficiency.

\begin{table}[tb] 
	\begin{tabular}{lcccccc}
		\toprule
		Kernel & Optimal $\varepsilon$ & $\Pj$ & $\rho_1$ & $E$ & Time (s) & $E$/Time$\times 10^5$\\
		
		\midrule
		RW&0.11&0.234&0.999&0.001&815&0.07\\
		MALA&0.48&0.574&0.939&0.029&1948&1.51\\
		\makecell[l]{Mirror (sparse)}
		&1&0.695&-0.011&0.395&5083&7.77\\
		\makecell[l]{MirrorMALA (sparse)}
		&1&0.865&-0.089&0.490&12726&3.85\\
		NUTS&Automatic&0.933&0.004&0.768&30532&2.52\\
		
		\bottomrule
	\end{tabular}
	\caption{The optimal size ($\varepsilon$), average acceptance probability ($\Pj$), lag-1 auto-correlation ($\rho_1$), efficiency ($E$), running time in seconds and efficiency per second for each of the RW, MALA, Mirror (sparse), MirrorMALA (sparse) and NUTS algorithms, which are used to sample from the posterior distribution of the GLMM for the polypharmacy data. Each of the $\rho_1$s and $E$s is an average over the 509 model parameters.} 
	\label{tab:polypharmacy}
\end{table}

\section{Discussion}

The Mirror kernel proposed by \citet{Thawornwattana2018} enables sampling on ``the other side'' of the target distribution, which is helpful to accelerate the mixing and improve the efficiency. However, the simple Mirror kernel is pretty sensitive to the  skewedness and dimensionality of the distribution. For example, the posterior of the Bayesian logistic regression model in Section 4.3 is just slightly skewed and has only 25 parameters, but the average acceptance probability of Mirror already drops to $1.5\%$. 

Therefore we generate the Mirror kernel to a wide class of Mirror-type kernels, by embedding the mirror reflection into various MH proposal distributions, many of which are more sophisticated than RW. In this paper, we focus on the MirrorMALA kernel, obtained by embedding the mirror reflection into the MALA kernel. MirrorMALA not only proposes around the mirror image of the current state, but also uses the gradient information to correct the location of the proposal, making it closer to the high-density region of the target distribution. MirrorMALA produces higher efficiency than both Mirror and MALA. More importantly, MirrorMALA is more robust to skewedness and dimensionality. For both the 100-D Gaussian and Bayesian logistic regression examples in Section 4, MirrorMALA exhibits high acceptance probability and high efficiency. 

MirrorMALA is easy to tune, but requires the gradients of the target distributions. If the gradients are difficult to calculate analytically, we can use numerical methods to approximate them. We experimented with a 2-D and 10-D normal distributions respectively, replaced the gradients with simple differentials in the MirrorMALA algorithm, and found that the efficiency is almost identical to MirrorMALA with the exact gradients. However, using simple differentials is computationally expensive since it evaluates the likelihood for many more times. Therefore, for high-dimensional problems we suggest to use more advanced approximation methods, such as automatic differentiation, sub-gradient, etc. 

As the examples in Sections~4 and 5 show, MirrorMALA can achieve similar performance as HMC, but takes much less time. Therefore, in many cases, MirroMALA can be a good substitute of HMC. However, high dimensionality may still have significant effect on the efficiency of MirrorMALA. In this paper, we recommend to use whitening transformation and block sampling to handle high-dimensional and complex distributions. For the GLMMs, we introduced a sparse whitening matrix which preserves the conditional independence of random effects, so that the block sampling does not lead to dramatic increase of computational time. In fact, the block sampling of random effects can easily be parallelized and thus the computational time can get further reduced. Although we simply use the GLMMs to illustrate the effects of sparse whitening and blocking, hierarchical models with conditionally independent structures are widely used in practice, so that the strategies we introduced in Section 5.4 can be easily applied to other scenarios.  

In the future, it would be interesting to investigate the effect of combining the Mirror-type kernels with the Riemann manifold. The whitening transformation only accounts for linear correlations between variables, while the Riemann manifold also describes other types of correlations \citep{girolami2011riemann}. Also, it can be  considered to develop the stochastic gradient version of MirrorMALA to deal with large datasets.

\backmatter

\bmhead{Supplementary information}

(1) The effect of $c$ on the efficiency of the Mirror and MirrorMALA kernels with $\epsilon = $0.25, 0.75, 1, 2. (2) Impact of the accuracy of the estimates for $\mu^*$ and $\Sigma^*$ on the efficiency of the algorithms. (3) Results of sampling from the $\N(0, I_d)$ distributions. (4) Marginal distributions of all the parameters in the Bayesian logistic regression model.\\
The R codes for the Bayesian logistic regression and Bayesian GLMM examples in this paper are available from the github repository: https://github.com/gen-evo-bayes/Mirror-type.

\bmhead{Acknowledgements}

This study has been supported by a National Natural Science Foundation of China grant (12101295), a Guangdong Natural Science Foundation
grant (2022A1515011767) and a Shenzhen outstanding youth grant (RCYX20221008093033012) to X.J.


\bibliographystyle{ba}
\bibliography{sample}


\begin{thebibliography}{23}
\ifx \bisbn   \undefined \def \bisbn  #1{ISBN #1}\fi
\ifx \binits  \undefined \def \binits#1{#1}\fi
\ifx \bauthor  \undefined \def \bauthor#1{#1}\fi
\ifx \batitle  \undefined \def \batitle#1{#1}\fi
\ifx \bjtitle  \undefined \def \bjtitle#1{#1}\fi
\ifx \bvolume  \undefined \def \bvolume#1{\textbf{#1}}\fi
\ifx \byear  \undefined \def \byear#1{#1}\fi
\ifx \bissue  \undefined \def \bissue#1{#1}\fi
\ifx \bfpage  \undefined \def \bfpage#1{#1}\fi
\ifx \blpage  \undefined \def \blpage #1{#1}\fi
\ifx \burl  \undefined \def \burl#1{\textsf{#1}}\fi
\ifx \doiurl  \undefined \def \doiurl#1{\url{https://doi.org/#1}}\fi
\ifx \betal  \undefined \def \betal{\textit{et al.}}\fi
\ifx \binstitute  \undefined \def \binstitute#1{#1}\fi
\ifx \binstitutionaled  \undefined \def \binstitutionaled#1{#1}\fi
\ifx \bctitle  \undefined \def \bctitle#1{#1}\fi
\ifx \beditor  \undefined \def \beditor#1{#1}\fi
\ifx \bpublisher  \undefined \def \bpublisher#1{#1}\fi
\ifx \bbtitle  \undefined \def \bbtitle#1{#1}\fi
\ifx \bedition  \undefined \def \bedition#1{#1}\fi
\ifx \bseriesno  \undefined \def \bseriesno#1{#1}\fi
\ifx \blocation  \undefined \def \blocation#1{#1}\fi
\ifx \bsertitle  \undefined \def \bsertitle#1{#1}\fi
\ifx \bsnm \undefined \def \bsnm#1{#1}\fi
\ifx \bsuffix \undefined \def \bsuffix#1{#1}\fi
\ifx \bparticle \undefined \def \bparticle#1{#1}\fi
\ifx \barticle \undefined \def \barticle#1{#1}\fi
\bibcommenthead
\ifx \bconfdate \undefined \def \bconfdate #1{#1}\fi
\ifx \botherref \undefined \def \botherref #1{#1}\fi
\ifx \url \undefined \def \url#1{\textsf{#1}}\fi
\ifx \bchapter \undefined \def \bchapter#1{#1}\fi
\ifx \bbook \undefined \def \bbook#1{#1}\fi
\ifx \bcomment \undefined \def \bcomment#1{#1}\fi
\ifx \oauthor \undefined \def \oauthor#1{#1}\fi
\ifx \citeauthoryear \undefined \def \citeauthoryear#1{#1}\fi
\ifx \endbibitem  \undefined \def \endbibitem {}\fi
\ifx \bconflocation  \undefined \def \bconflocation#1{#1}\fi
\ifx \arxivurl  \undefined \def \arxivurl#1{\textsf{#1}}\fi
\csname PreBibitemsHook\endcsname

\bibitem[\protect\citeauthoryear{Tan and Nott}{2018}]{tan2018gaussian}
\begin{barticle}
\bauthor{\bsnm{Tan}, \binits{L.S.}},
\bauthor{\bsnm{Nott}, \binits{D.J.}}:
\batitle{Gaussian variational approximation with sparse precision matrices}.
\bjtitle{Statistics and Computing}
\bvolume{28},
\bfpage{259}--\blpage{275}
(\byear{2018})
\end{barticle}
\endbibitem

\bibitem[\protect\citeauthoryear{Metropolis et~al.}{1953}]{Metropolis1953}
\begin{barticle}
\bauthor{\bsnm{Metropolis}, \binits{N.}},
\bauthor{\bsnm{Rosenbluth}, \binits{A.W.}},
\bauthor{\bsnm{Rosenbluth}, \binits{M.N.}},
\bauthor{\bsnm{Teller}, \binits{A.H.}},
\bauthor{\bsnm{Teller}, \binits{E.}}:
\batitle{Equation of state calculations by fast computing machines}.
\bjtitle{J. Chem. Phys.}
\bvolume{21},
\bfpage{1087}--\blpage{1092}
(\byear{1953})
\end{barticle}
\endbibitem

\bibitem[\protect\citeauthoryear{Hastings}{1970}]{Hastings1970}
\begin{barticle}
\bauthor{\bsnm{Hastings}, \binits{W.K.}}:
\batitle{{M}onte {C}arlo sampling methods using {M}arkov chains and their applications}.
\bjtitle{Biometrika}
\bvolume{57},
\bfpage{97}--\blpage{109}
(\byear{1970})
\end{barticle}
\endbibitem

\bibitem[\protect\citeauthoryear{Robert and Casella}{2004}]{robert2004monte}
\begin{botherref}
\oauthor{\bsnm{Robert}, \binits{C.P.}},
\oauthor{\bsnm{Casella}, \binits{G.}}:
Monte carlo statistical methods.
Springer Texts in Statistics,
274
(2004)
\end{botherref}
\endbibitem

\bibitem[\protect\citeauthoryear{Ross}{2014}]{ross2014introduction}
\begin{bbook}
\bauthor{\bsnm{Ross}, \binits{S.M.}}:
\bbtitle{Introduction to Probability Models}.
\bpublisher{Academic press},
\blocation{Cambridge, MA}
(\byear{2014})
\end{bbook}
\endbibitem

\bibitem[\protect\citeauthoryear{Roberts et~al.}{1997}]{roberts1997}
\begin{barticle}
\bauthor{\bsnm{Roberts}, \binits{G.O.}},
\bauthor{\bsnm{Gelman}, \binits{A.}},
\bauthor{\bsnm{Gilks}, \binits{W.R.}}:
\batitle{Weak convergence and optimal scaling of random walk {M}etropolis algorithms}.
\bjtitle{The Annals of Applied Probability}
\bvolume{7},
\bfpage{110}--\blpage{120}
(\byear{1997})
\end{barticle}
\endbibitem

\bibitem[\protect\citeauthoryear{Gelman et~al.}{1996}]{gelman1996efficient}
\begin{barticle}
\bauthor{\bsnm{Gelman}, \binits{A.}},
\bauthor{\bsnm{Roberts}, \binits{G.O.}},
\bauthor{\bsnm{Gilks}, \binits{W.R.}}:
\batitle{Efficient metropolis jumping rules}.
\bjtitle{Bayesian statistics}
\bvolume{5},
\bfpage{599}--\blpage{608}
(\byear{1996})
\end{barticle}
\endbibitem

\bibitem[\protect\citeauthoryear{Yang and Rodríguez}{2013}]{Yang2013}
\begin{barticle}
\bauthor{\bsnm{Yang}, \binits{Z.}},
\bauthor{\bsnm{Rodríguez}, \binits{C.E.}}:
\batitle{Searching for efficient markov chain monte carlo proposal kernels}.
\bjtitle{Proc. Natl. Acad. Sci. U.S.A.}
\bvolume{110}(\bissue{48}),
\bfpage{19307}--\blpage{19312}
(\byear{2013})
\end{barticle}
\endbibitem

\bibitem[\protect\citeauthoryear{Jiao et~al.}{2025}]{Jiao2025}
\begin{botherref}
\oauthor{\bsnm{Jiao}, \binits{X.}},
\oauthor{\bsnm{Huang}, \binits{J.}},
\oauthor{\bsnm{Yang}, \binits{Z.}}:
Super-efficient markov chain monte carlo algorithms for bayesian inference in population genomics.
Statistical Applications in Genetics and Molecular Biology
\textbf{submitted}
(2025)
\end{botherref}
\endbibitem

\bibitem[\protect\citeauthoryear{Thawornwattana et~al.}{2018}]{Thawornwattana2018}
\begin{barticle}
\bauthor{\bsnm{Thawornwattana}, \binits{Y.}},
\bauthor{\bsnm{Dalquen}, \binits{D.A.}},
\bauthor{\bsnm{Yang}, \binits{Z.}}:
\batitle{Coalescent analysis of phylogenomic data confidently resolves the species relationships in the \textit{Anopheles gambiae} species complex}.
\bjtitle{Mol. Biol. Evol.}
\bvolume{35}(\bissue{10}),
\bfpage{2512}--\blpage{2527}
(\byear{2018})
\end{barticle}
\endbibitem

\bibitem[\protect\citeauthoryear{Roberts and Tweedie}{1996}]{roberts1996exponential}
\begin{botherref}
\oauthor{\bsnm{Roberts}, \binits{G.O.}},
\oauthor{\bsnm{Tweedie}, \binits{R.L.}}:
Exponential convergence of langevin distributions and their discrete approximations.
Bernoulli,
341--363
(1996)
\end{botherref}
\endbibitem

\bibitem[\protect\citeauthoryear{Roberts and Rosenthal}{1998}]{roberts1998optimal}
\begin{barticle}
\bauthor{\bsnm{Roberts}, \binits{G.O.}},
\bauthor{\bsnm{Rosenthal}, \binits{J.S.}}:
\batitle{Optimal scaling of discrete approximations to langevin diffusions}.
\bjtitle{Journal of the Royal Statistical Society: Series B (Statistical Methodology)}
\bvolume{60}(\bissue{1}),
\bfpage{255}--\blpage{268}
(\byear{1998})
\end{barticle}
\endbibitem

\bibitem[\protect\citeauthoryear{Duane et~al.}{1987}]{duane1987hybrid}
\begin{barticle}
\bauthor{\bsnm{Duane}, \binits{S.}},
\bauthor{\bsnm{Kennedy}, \binits{A.D.}},
\bauthor{\bsnm{Pendleton}, \binits{B.J.}},
\bauthor{\bsnm{Roweth}, \binits{D.}}:
\batitle{Hybrid monte carlo}.
\bjtitle{Physics letters B}
\bvolume{195}(\bissue{2}),
\bfpage{216}--\blpage{222}
(\byear{1987})
\end{barticle}
\endbibitem

\bibitem[\protect\citeauthoryear{Neal}{2011}]{neal2011mcmc}
\begin{barticle}
\bauthor{\bsnm{Neal}, \binits{R.M.}}:
\batitle{Mcmc using hamiltonian dynamics}.
\bjtitle{Handbook of markov chain monte carlo}
\bvolume{2}(\bissue{11}),
\bfpage{2}
(\byear{2011})
\end{barticle}
\endbibitem

\bibitem[\protect\citeauthoryear{Beskos et~al.}{2013}]{beskos2013hmc}
\begin{barticle}
\bauthor{\bsnm{Beskos}, \binits{A.}},
\bauthor{\bsnm{Pillai}, \binits{N.}},
\bauthor{\bsnm{Roberts}, \binits{G.}},
\bauthor{\bsnm{Sanz-Serna}, \binits{J.-M.}},
\bauthor{\bsnm{Stuart}, \binits{A.}}:
\batitle{Optimal tuning of the hybrid {M}onte {C}arlo algorithm}.
\bjtitle{Bernoulli}
\bvolume{19}(\bissue{5A}),
\bfpage{1501}--\blpage{1534}
(\byear{2013})
\end{barticle}
\endbibitem

\bibitem[\protect\citeauthoryear{Hoffman et~al.}{2014}]{hoffman2014no}
\begin{barticle}
\bauthor{\bsnm{Hoffman}, \binits{M.D.}},
\bauthor{\bsnm{Gelman}, \binits{A.}}, \betal:
\batitle{The no-u-turn sampler: adaptively setting path lengths in hamiltonian monte carlo.}
\bjtitle{J. Mach. Learn. Res.}
\bvolume{15}(\bissue{1}),
\bfpage{1593}--\blpage{1623}
(\byear{2014})
\end{barticle}
\endbibitem

\bibitem[\protect\citeauthoryear{Girolami and Calderhead}{2011}]{girolami2011riemann}
\begin{barticle}
\bauthor{\bsnm{Girolami}, \binits{M.}},
\bauthor{\bsnm{Calderhead}, \binits{B.}}:
\batitle{Riemann manifold langevin and hamiltonian monte carlo methods}.
\bjtitle{Journal of the Royal Statistical Society Series B: Statistical Methodology}
\bvolume{73}(\bissue{2}),
\bfpage{123}--\blpage{214}
(\byear{2011})
\end{barticle}
\endbibitem

\bibitem[\protect\citeauthoryear{Roberts and Stramer}{2002}]{roberts2002langevin}
\begin{barticle}
\bauthor{\bsnm{Roberts}, \binits{G.O.}},
\bauthor{\bsnm{Stramer}, \binits{O.}}:
\batitle{Langevin diffusions and metropolis-hastings algorithms}.
\bjtitle{Methodology and computing in applied probability}
\bvolume{4},
\bfpage{337}--\blpage{357}
(\byear{2002})
\end{barticle}
\endbibitem

\bibitem[\protect\citeauthoryear{Frank and Asuncion}{2010}]{FrankAsuncion2010}
\begin{botherref}
\oauthor{\bsnm{Frank}, \binits{A.}},
\oauthor{\bsnm{Asuncion}, \binits{A.}}:
Uci machine learning repository.
Statistical Applications in Genetics and Molecular Biology
(2010)
\end{botherref}
\endbibitem

\bibitem[\protect\citeauthoryear{Jiang and Nguyen}{2007}]{jiang2007linear}
\begin{bbook}
\bauthor{\bsnm{Jiang}, \binits{J.}},
\bauthor{\bsnm{Nguyen}, \binits{T.}}:
\bbtitle{Linear and Generalized Linear Mixed Models and Their Applications}
vol. \bseriesno{1}.
\bpublisher{Springer},
\blocation{New York, NY}
(\byear{2007})
\end{bbook}
\endbibitem

\bibitem[\protect\citeauthoryear{Thall and Vail}{1990}]{thall1990some}
\begin{botherref}
\oauthor{\bsnm{Thall}, \binits{P.F.}},
\oauthor{\bsnm{Vail}, \binits{S.C.}}:
Some covariance models for longitudinal count data with overdispersion.
Biometrics,
657--671
(1990)
\end{botherref}
\endbibitem

\bibitem[\protect\citeauthoryear{Breslow and Clayton}{1993}]{breslow1993approximate}
\begin{barticle}
\bauthor{\bsnm{Breslow}, \binits{N.E.}},
\bauthor{\bsnm{Clayton}, \binits{D.G.}}:
\batitle{Approximate inference in generalized linear mixed models}.
\bjtitle{Journal of the American statistical Association}
\bvolume{88}(\bissue{421}),
\bfpage{9}--\blpage{25}
(\byear{1993})
\end{barticle}
\endbibitem

\bibitem[\protect\citeauthoryear{Hosmer~Jr et~al.}{2013}]{hosmer2013applied}
\begin{bbook}
\bauthor{\bsnm{Hosmer~Jr}, \binits{D.W.}},
\bauthor{\bsnm{Lemeshow}, \binits{S.}},
\bauthor{\bsnm{Sturdivant}, \binits{R.X.}}:
\bbtitle{Applied Logistic Regression}.
\bpublisher{John Wiley \& Sons},
\blocation{Hoboken, NJ}
(\byear{2013})
\end{bbook}
\endbibitem

\end{thebibliography}

\end{document}


\textbf{\Large Supplementary Material}
\section{The effect of $c$ on the efficiency of the Mirror and MirrorMALA kernels with $\epsilon = 0.25, 0.75, 1, 2$}

\begin{figure}[!htbp] 
	\begin{center}
		\includegraphics[scale = 0.6]{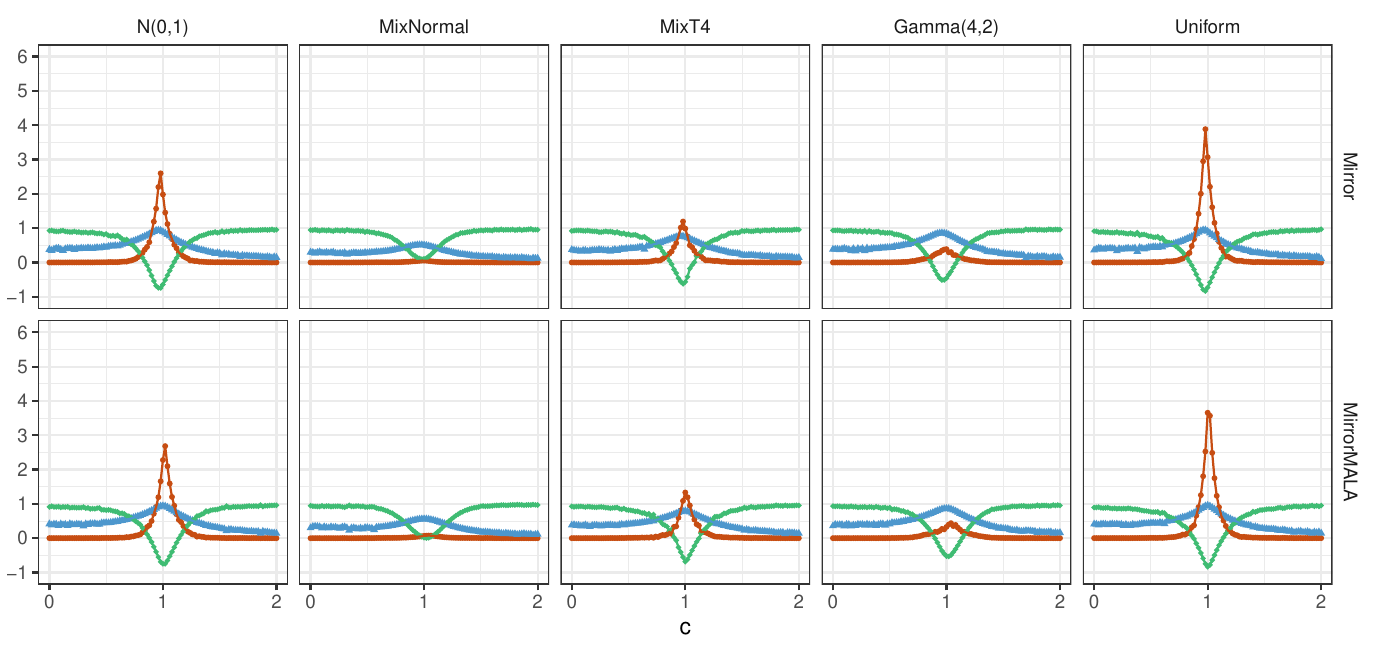}
	\end{center}
    \vspace{-12pt}
	\caption[]{$\Pj$ (blue triangle), $\rho_{1}$ (green square) and $E$ (red dot) of Mirror and MirrorMALA for sampling from the five 1-D targets in figure 3 of the main text. Here $\epsilon = 0.25$.}
	\label{fig:pjump-rho1-eff-c-e-025}
\end{figure}

\begin{figure}[!htbp] 
	\begin{center}
		\includegraphics[scale = 0.6]{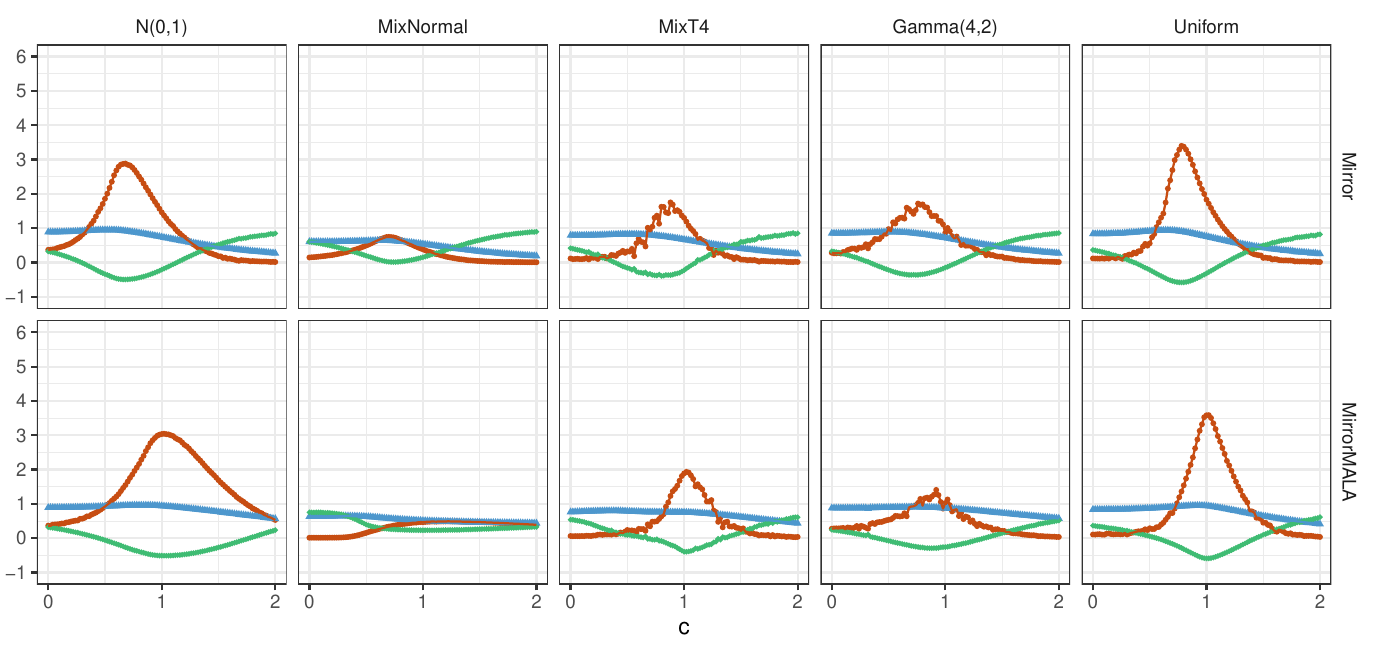}
	\end{center}
    \vspace{-12pt}
	\caption[]{$\Pj$, $\rho_{1}$ and $E$ of MALA and MirrorMALA for sampling from the five 1-D targets. Here $\epsilon = 0.75$.}
	\label{fig:pjump-rho1-eff-c-e-075}
\end{figure}

\begin{figure}[!htbp] 
	\begin{center}
		\includegraphics[scale = 0.6]{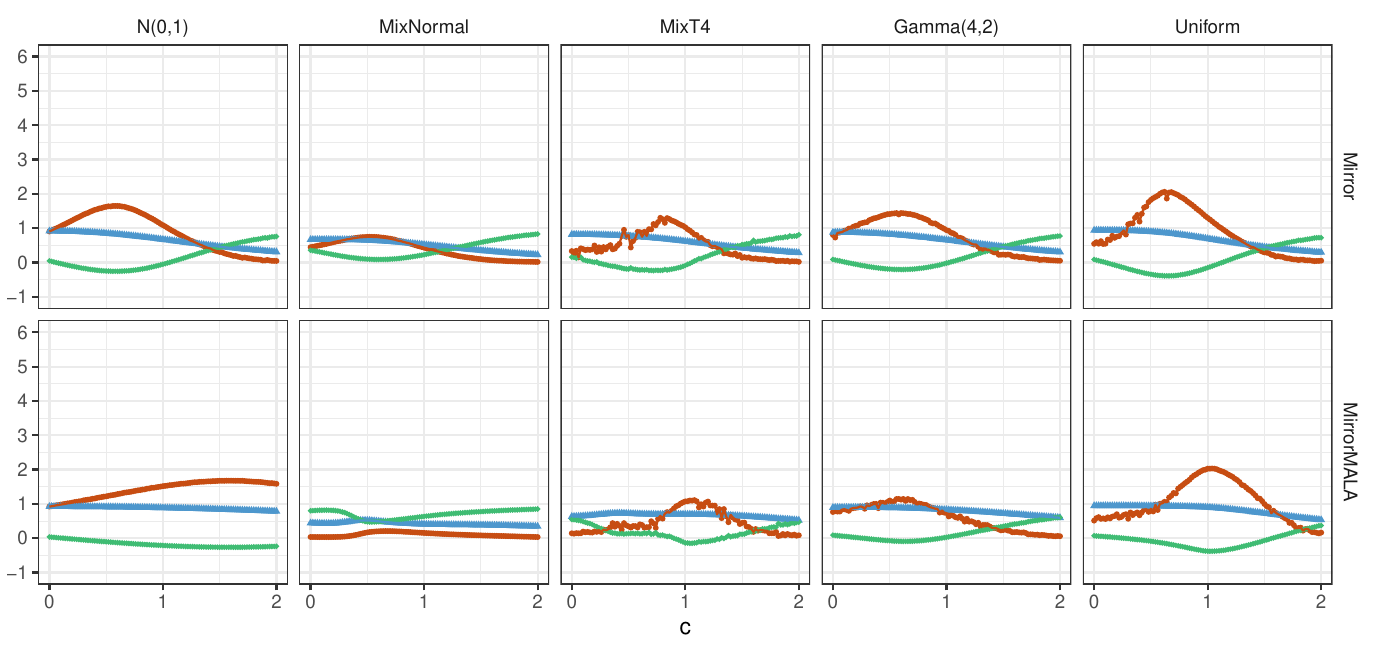}
	\end{center}
    \vspace{-12pt}
    \caption[]{$\Pj$, $\rho_{1}$ and $E$ of MALA and MirrorMALA for sampling from the five 1-D targets. Here $\epsilon = 1$.}
	\label{fig:pjump-rho1-eff-c-e-100}
\end{figure}

\begin{figure}[!htbp] 
	\begin{center}
		\includegraphics[scale = 0.6]{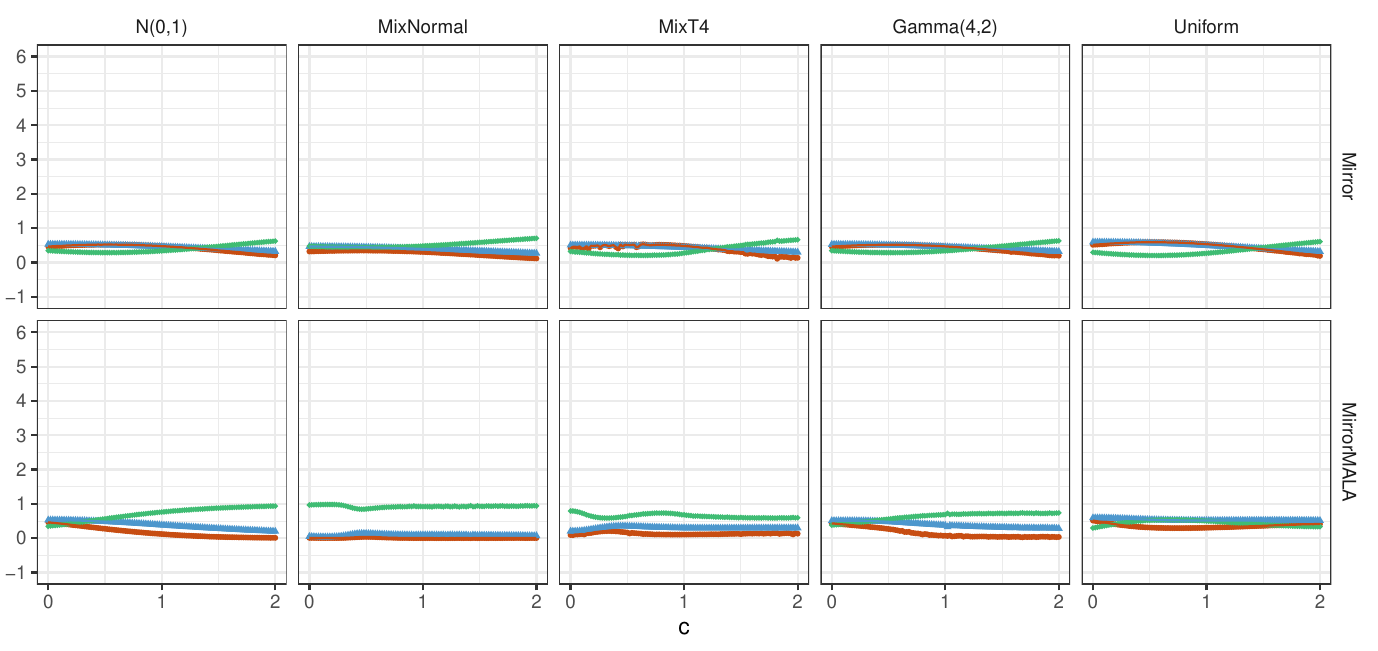}
	\end{center}
	\vspace{-12pt}
	\caption[]{$\Pj$, $\rho_{1}$ and $E$ of MALA and MirrorMALA for sampling from the five 1-D targets. Here $\epsilon = 2$.}
	\label{fig:pjump-rho1-eff-c-e-200}
\end{figure}

\section{Impact of the accuracy of the estimates for $\mu^*$ and $\Sigma^*$ on the efficiency of the algorithms}

\begin{figure}[h!] 
	\begin{center}
		\includegraphics[scale = 0.58]{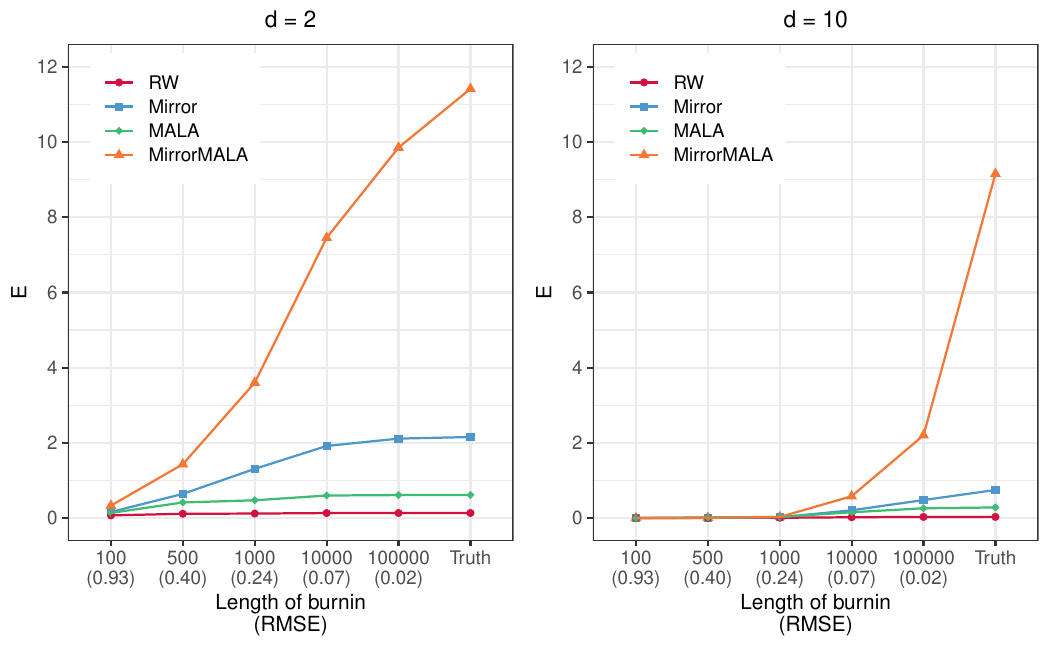}
	\end{center}
	\caption[]{The efficiency ($E$) plotted against the length of the burn-in for the RW, Mirror, MALA, and MirrorMALA algorithms sampling from the 2-D and 10-D Gaussian distributions described in Section~4 of the main text. The RMSE in parentheses under each length of burn-in is the square root of the average squared distance between the true values of the mean and variance-covariance parameters in the target and their estimates calculated using the burn-in.}
	\label{fig:binormal-results}
\end{figure}

\section{Results of sampling from the $\N(0, I_d)$ distributions}
\begin{table}[h!]
	\begin{tabular}{ccccccc}
		\toprule
		$d$ & RW & Mirror1/2 & Mirror1 & MALA & MirrorMALA1/2 & MirrorMALA1\\
		
		\midrule
		2& 0.352& 0.757& 0.552& 0.651& 0.968& 0.876\\
		3& 0.315& 0.693& 0.452& 0.639& 0.954& 0.839\\
		4& 0.279& 0.640& 0.371& 0.612& 0.944& 0.812\\
		5& 0.274& 0.600& 0.315& 0.600& 0.936& 0.784\\
		6& 0.267& 0.559& 0.262& 0.600& 0.915& 0.765\\
		7& 0.260& 0.522& 0.229& 0.604& 0.907& 0.750\\
		8& 0.254& 0.490& 0.192& 0.601& 0.895& 0.726\\
		9& 0.260& 0.462& 0.167& 0.592& 0.883& 0.707\\
		10& 0.267& 0.435& 0.143& 0.591& 0.880& 0.692\\
		
		\bottomrule
	\end{tabular}
	\caption{The average acceptance probability ($\Pj$) of the RW, Mirror1/2 (Mirror with $\epsilon = 0.5$), Mirror1 (Mirror with $\epsilon = 1$), MALA, MirrorMALA1/2 (MirrorMALA with $\epsilon = 0.5$), and MirrorMALA1 (MirrorMALA with $\epsilon = 1$) kernels respectively for sampling from the $\N(0, I_d)$ target distributions with dimension $d=2, 3,\dots, 10$.} 
	\label{table:md_ap}
\end{table}

\begin{table}[h!]
	\begin{tabular}{ccccccc}
		\toprule
		$d$ & RW & Mirror1/2 & Mirror1 & MALA & MirrorMALA1/2 & MirrorMALA1\\
		
		\midrule
		2& 0.762& -0.366& 0.066& 0.218& -0.794& -0.327\\
		3& 0.825& -0.274& 0.210& 0.309& -0.772& -0.278\\
		4& 0.862& -0.189& 0.338& 0.367& -0.755& -0.246\\
		5& 0.886& -0.124& 0.431& 0.413& -0.740& -0.206\\
		6& 0.902& -0.055& 0.520& 0.450& -0.698& -0.184\\
		7& 0.915&  0.010& 0.578& 0.481& -0.687& -0.168\\
		8& 0.925&  0.067& 0.643& 0.505& -0.663& -0.129\\
		9& 0.932&  0.119& 0.687& 0.525& -0.642& -0.100\\
		10& 0.939&  0.167& 0.731& 0.543& -0.639& -0.079\\
		
		\bottomrule
	\end{tabular}
	\caption{The lag-1 autocorrelation ($\rho_1$) of the RW, Mirror1/2, Mirror1, MALA, MirrorMALA1/2 and MirrorMALA1 kernels respectively for sampling from $\N(0, I_d)$ with $d=2, 3,\dots, 10$. Each of the $\rho_1$s is an averages over $d$ variables.} 
	\label{table:md_ac}
\end{table}

\begin{table}[h!]
	\begin{tabular}{ccccccc}
		\toprule
		$d$ & RW & Mirror1/2 & Mirror1 & MALA & MirrorMALA1/2 & MirrorMALA1\\
		
		\midrule
		2& 0.135& 2.104& 0.872& 0.616& 8.261& 1.945\\
		3& 0.096& 1.658& 0.641& 0.506& 6.895& 1.730\\
		4& 0.074& 1.331& 0.477& 0.439& 6.189& 1.600\\
		5& 0.061& 1.109& 0.376& 0.394& 5.398& 1.450\\
		6& 0.051& 0.981& 0.293& 0.360& 4.013& 1.370\\
		7& 0.044& 0.809& 0.246& 0.334& 3.363& 1.312\\
		8& 0.039& 0.676& 0.193& 0.313& 3.021& 1.194\\
		9& 0.035& 0.578& 0.162& 0.296& 2.622& 1.115\\
		10& 0.032& 0.517& 0.133& 0.281& 2.487& 1.052\\
		
		\bottomrule
	\end{tabular}
	\caption{The efficiency ($E$) of the RW, Mirror1/2, Mirror1, MALA, MirrorMALA1/2 and MirrorMALA1 kernels respectively for sampling from $\N(0, I_d)$ with $d=2, 3,\dots, 10$. Each of the $\rho_1$s is an averages over $d$ variables.} 
	\label{table:md_eff}
\end{table}

\begin{table}[h!]
	\begin{tabular}{ccccccc}
		\toprule
		$d$ & RW & Mirror1/2 & Mirror1 & MALA & MirrorMALA1/2 & MirrorMALA1\\
		
		\midrule
		2&5.5&5.6&5.7&6.9&6.9&6.7\\
		3&5.7&5.8&5.8&7.2&6.9&7.0\\
		4&5.9&6.0&5.9&7.2&7.1&7.0\\
		5&5.9&6.2&6.1&7.3&7.3&7.3\\
		6&6.0&6.2&6.3&7.3&7.3&7.3\\
		7&6.2&6.2&6.3&7.4&7.3&8.7\\
		8&6.3&6.5&6.5&7.5&7.4&7.6\\
		9&6.3&6.6&6.7&7.8&7.7&7.9\\
		10&6.4&6.6&6.7&7.8&7.8&7.9\\
		
		\bottomrule
	\end{tabular}
	\caption{The running time in seconds of the RW, Mirror1/2, Mirror1, MALA, MirrorMALA1/2 and MirrorMALA1 kernels respectively for sampling from $\N(0, I_d)$ with $d=2, 3,\dots, 10$.} 
	\label{table:md_rt}
\end{table}

\begin{table}[h!]
	\begin{tabular}{ccccccc}
		\toprule
		$d$ & RW & Mirror1/2 & Mirror1 & MALA & MirrorMALA1/2 & MirrorMALA1\\
		
		\midrule
		2& 0.025& 0.374& 0.152& 0.089& 1.205& 0.289\\
		3& 0.017& 0.288& 0.111& 0.071& 0.993& 0.249\\
		4& 0.013& 0.223& 0.080& 0.061& 0.872& 0.227\\
		5& 0.010& 0.180& 0.061& 0.054& 0.744& 0.199\\
		6& 0.009& 0.157& 0.047& 0.049& 0.549& 0.188\\
		7& 0.007& 0.130& 0.039& 0.045& 0.462& 0.150\\
		8& 0.006& 0.104& 0.030& 0.042& 0.407& 0.157\\
		9& 0.006& 0.088& 0.024& 0.038& 0.339& 0.141\\
		10& 0.005& 0.078& 0.020& 0.036& 0.319& 0.134\\
		
		\bottomrule
	\end{tabular}
	\caption{The efficiency per second of the RW, Mirror1/2, Mirror1, MALA, MirrorMALA1/2 and MirrorMALA1 kernels respectively for sampling from $\N(0, I_d)$ with $d=2, 3,\dots, 10$.} 
	\label{table:md_effptu}
\end{table}

\section{Marginal distributions of all the parameters in the Bayesian logistic regression model}

\begin{figure}[h!] 
	\begin{center}
		\includegraphics[scale = 0.42]{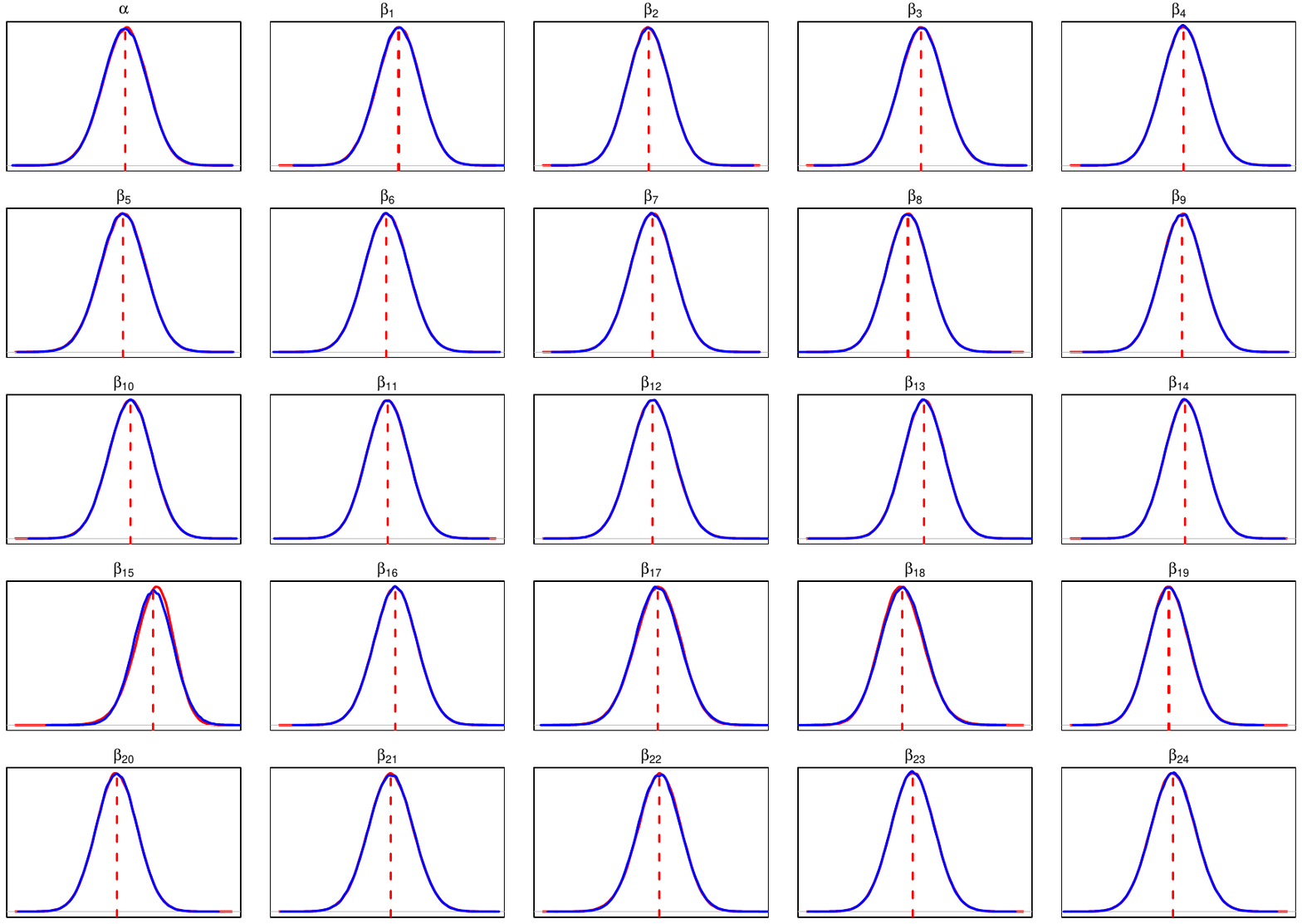}
	\end{center}
	\caption[]{Marginal distributions (red) of all the parameters in the Bayesian logistic regression model described in Section~4.3 of the main text. The densities are estimated using the samples generated by the MirrorMALA1/2 algorithm, and the blue lines are the normal densities with the same mean and variance as the parameters.}
	\label{fig:logistic-marginal}
\end{figure}